%% file: GAII_v2.tex
\documentclass[a4paper,fleqn,usenatbib]{mnras}

\usepackage{newtxtext,newtxmath}
\usepackage[T1]{fontenc}
\usepackage{ae,aecompl}

\usepackage{graphicx}	
\usepackage{amsmath}	
\usepackage{amssymb}	

\usepackage{makecell}
\usepackage{lscape}
\usepackage{color}
\usepackage[figuresright]{rotating}
\usepackage{bbm}

\def\mean#1{\left< #1 \right>}

\definecolor{green2}{rgb}{0.13, 0.55, 0.13}
\definecolor{gray2}{rgb}{0.6, 0.6, 0.6}
\definecolor{orange}{rgb}{1.0, 0.49, 0}
\definecolor{blue2}{rgb}{0.25,0.5,1}

\title[Satellite alignments]{Intrinsic Alignment in redMaPPer clusters -- II. Radial alignment of satellites toward cluster centers}

\author[Huang et al.]{
Hung-Jin Huang$^{1}$\thanks{E-mail: hungjinh@andrew.cmu.edu},
Rachel Mandelbaum$^{1}$\thanks{E-mail: rmandelb@andrew.cmu.edu},
Peter E. Freeman$^{1, 2}$, 
Yen-Chi Chen$^{3}$, \newauthor
\ Eduardo Rozo$^{4} \&$ 
Eli Rykoff$^{5}$ 
\vspace{0.1in}\\
$^{1}$McWilliams Center for Cosmology, Department of Physics, Carnegie Mellon University, Pittsburgh, PA 15213, USA\\
$^{2}$Department of Statistics, Carnegie Mellon University, Pittsburgh, PA 15213, USA\\
$^{3}$Department of Statistics, University of Washington, Seattle, WA 98195, USA\\
$^{4}$Department of Physics, University of Arizona, 1118 E. Fourth St., Tucson, AZ 85721, USA\\
$^{5}$SLAC National Accelerator Laboratory, Menlo Park, CA 94025, USA \\
}

\date{Accepted XXX. Received YYY; in original form ZZZ}

\pubyear{2016}

\begin{document}
\label{firstpage}
\pagerange{\pageref{firstpage}--\pageref{lastpage}}
\maketitle

\begin{abstract}
We study the orientations of satellite galaxies in redMaPPer clusters constructed from the Sloan Digital Sky Survey at $0.1<z<0.35$ to determine whether there is any preferential tendency for satellites to point radially toward cluster centers.
We analyze the satellite alignment (SA) signal based on three shape measurement methods (re-Gaussianization, de Vaucouleurs, and isophotal shapes), which trace galaxy light profiles at different radii.
The measured SA signal depends on these shape measurement methods.
We detect the strongest SA signal in isophotal shapes, followed by de Vaucouleurs shapes.
While no net SA signal is detected using re-Gaussianization shapes across the entire sample, the
observed SA signal reaches a statistically significant level
when limiting to a subsample of higher luminosity satellites.
We further investigate the impact of noise, systematics, and real physical isophotal twisting effects 
in the comparison between the SA signal detected via different shape measurement methods.
Unlike previous studies, which only consider the dependence of SA on a few parameters, here
  we explore a total of 17 galaxy and cluster properties, using a statistical model averaging
  technique to naturally account for parameter correlations and identify significant SA predictors.
We find that the measured SA signal is strongest for satellites with the following characteristics: higher luminosity, smaller distance to the cluster center, rounder in shape, higher bulge fraction, and distributed preferentially along the major axis directions of their centrals.
Finally, we provide physical explanations for the identified dependences, and discuss the connection to theories of SA.

\end{abstract}



\input{1_introduction}

\input{2_data}

\input{3_SAsignal}

\input{4_LR}

\input{5_cpShapes}

\input{6_discussion}

\input{7_summary}


\section*{Acknowledgements}

\input{ack}



\bibliographystyle{mnras}
\bibliography{GAII_reference} 



\appendix

\input{8_appendix}

\input{9_appendix2}


\bsp	
\label{lastpage}
\end{document}

%% file: 1_introduction.tex

\section{Introduction}
\label{sec:intro}


The projected orientations of galaxies observed on sky are not random, but rather exhibit some coherent patterns related to the matter distribution in the Universe. 
Galaxy shapes tend to point towards overdense regions, leaving a net preference of correlated orientations. This phenomenon, known as ``intrinsic'' alignments (IA), contains important information about structure formation and galaxy evolution (for recent reviews, see \citealt{Joachimi15,Kirk15,Kiessling15}).
Besides the physically-induced alignment signal, the images of galaxies located behind overdense
structures tend to be distorted tangentially with respect to those structures, producing the
apparent tangential alignment signal that is the key characteristic of gravitational lensing. This
lensing effect is used as a tool to map the distribution of dark matter in the Universe, to study the growth of structure, and to constrain cosmological parameters (see e.g.\ \citealt{Massey10,Weinberg13,Mandelbaum13}). 
The presence of IA challenges the process of interpreting the observed shape correlations
(intrinsic$+$apparent) in terms of the basic physics that generates lensing signals.
Ongoing surveys such as the Dark Energy Survey (DES, \citealt{DES16}), the Kilo-Degree Survey (KiDS, \citealt{deJong15}), Hyper Suprime-Cam Survey (HSC, \citealt{Miyazaki12}), 
and future surveys like the Large Synoptic Survey Telescope (LSST, \citealt{LSST09}), Euclid
\citep{Laureijs11}, and the Wide Field Infrared Survey Telescope (WFIRST, \citealt{Spergel15}) aim to constrain the cosmological constants to sub-percent precision, which requires precise removal of all possible systematics including
intrinsic alignments \citep[e.g.,][]{Blazek12, Krause16}.  
Quantifying the strength of IA signal and developing models that will enable its removal thus becomes one of the key steps to reach this goal.

IA have been detected over a wide range of scales. 
On scales above several Mpcs, red dispersion-dominated galaxies preferentially point towards overdense regions (see e.g.,
\citealt{Mandelbaum06, Hirata07, Okumura09, Joachimi11, Singh15} from the observational side, and
\citealt{Tenneti15, Chisari15} from numerical simulation). 
Part of this observed correlation originates from the tendency of galaxies to align towards
overdensities and with filamentary structures (see e.g., observation: \citealt{Zhang13,Tempel15}, and simulation: \citealt{Chen15b}). 
For red galaxies located in sheets, \citet{Zhang13} observed that they tend to have their major axes aligned parallel to the plane of the sheets.
For blue angular momentum-dominated galaxies, there is no significant detection of alignment so far \citep{Mandelbaum11}.
Besides the alignment of galaxies, people also found alignment between the shape of clusters with respect to the underlying density field (observation: \citealt{Smargon12, Uitert17}; simulation: \citealt{Hopkins05}).

The other alignment at intra-halo scale is satellite alignment, i.e.\ the preference of satellites to
align radially toward the cluster center.
The SA signal is relatively subtle compared with the strength of central galaxy alignment; along
with the difficulty of achieving accurate shape measurements on faint satellites whose light
profiles are more subject to contamination from neighboring galaxies, many conflicting observational results have been published.
Earlier works based on SDSS isophotal shape measurement, which trace the very outer part of the
galaxy light profiles, have reported detections of SA signal \citep{Pereira05, Agustsson06,
  Faltenbacher07}. However, later studies claimed that when using de Vaucouleurs shape, which puts
relatively more weight on the galaxy inner light profiles, satellite orientations are consistent
with random \citep{Siverd09, Hao10}. Studies that used shape measurements that are optimized for
lensing, which requires  corrections for many observational systematics,  reported non-detection of
satellite alignments \citep{Schneider13, Sifon15}. 
There is therefore some tension between past measurements, and reconciliation of that tension may
require investigation into the different galaxy populations used for these measurements and/or false
SA signals generated by systematics in isophotal shape measurements \citep[e.g.,][]{Hao10}.

Our current theoretical understanding of IA for red dispersion-dominated galaxies is that their orientations are affected by the tidal field of
the surrounding environment.  
On large scales, the linear alignment model \citep{Catelan01, Hirata04a} suggests that the shapes of
proto-galaxies are largely set by the primordial tidal field at their formation time, so that their
shape correlation with the matter field is frozen in since then and simply grows with the matter
power spectrum. 
The primordial tidal field also leaves its imprint on the assembly history of clusters by channeling
the majority of satellites into clusters through accretion along filamentary structures, which results in the observed cluster alignment phenomenon \citep{Hopkins05}. 
At small scales, the orientation of cluster central galaxies would also be generated by the same primordial tidal field, leaving the observed central galaxy alignment (see discussions in Paper I). 
While later non-linear evolutionary processes such as mergers or baryonic feedback from galaxies may
erase the alignment signals set by primordial tidal fields \citep{Tenneti17}, the late-time re-arranged structures can
set up new tidal environments that gradually torque galaxies to align \citep{Ciotti94, Kuhlen07, Pereira08,
  Faltenbacher08}. 
The timescales for tidal locking of satellites under the cluster potential depend on the eccentricity of infalling orbits as well as properties of satellites (e.g. angular momentum, morphology). As shown in the simulations of \citet{Pereira10}, within the time of one orbital period ($\sim$ 5 Gyr), a triaxial DM subhalo orbiting in circular orbit around a cluster potential becomes tidally locked, and it takes a lag of $\lesssim$2 Gyr, depending on the initial conditions, for the stellar components to respond.

In this work, we carry out SA measurements using re-Gaussianization, de Vaucouleurs and isophotal shapes that differ in sensitivity to the outskirt of a galaxy's light profile.
The size of the redMaPPer cluster catalogue provides the necessary statistical power to contrain SA signals at halo masses $\gtrsim 10^{14} M_\odot \rm\ h^{-1}$. 
The two main questions we aim to address in this paper are 
1) What causes the detected discrepancies in SA signals using different shape measurement methods?
2) Which satellite properties associated with which central galaxy and cluster properties are correlated with stronger SA signals?
We estimate the level of possible noises and systematics that could cause the inconsistent galaxy position angle (PA) measurements. As in paper I, we explore a large parameter pool which contains characteristic satellite, central galaxy, and cluster properties to identify important predictors of SA effect using linear regression analysis.

The paper is organized as follows. In Sec.~\ref{sec:data}, we describe our data and definitions of the physical parameters involved in the analysis. 
Sec.~\ref{sec:sat} presents the overall signal of SA alignment measured in redMaPPer clusters. 
Details of the linear regression and variable selection results are described in Sec.~\ref{sec:LR}. 
Sec.~\ref{sec:cp_shape} explores possible factors that cause the discrepancy in the measured SA
angle using three different shape measurement methods, and provides estimates of the degree of contribution from each factor. 
The physical origins of our identified featured predictors on the SA effect are discussed in Sec.~\ref{sec:discussion}.
We conclude and summarize our key findings in Sec.~\ref{sec:summary}.

Throughout this paper, we adopt the standard flat $\Lambda$CDM cosmology with $\Omega_m=0.3$ and
$\Omega_{\Lambda}=0.7$. All length and magnitude units use $H_0= 100$~km~s$^{-1}$~Mpc$^{-1}$.  
We use $\log$ as shorthand for the 10-based logarithm.


%% file: 2_data.tex

\section{Data and Measurements}
\label{sec:data}
All data used in this paper come from the SDSS \citep{York2000} surveys.
Here we describe the catalogs involved in our analysis, sample construction, 
and definitions of the cluster- and galaxy-related parameters. 
Most of the data and parameters remain the same as in Paper I, although some small differences
exist, as we will highlight in the relevant subsections below.  
In order to properly interpret the measured satellite alignments, Paper II puts more focus on exploring systematics in the different shape measurement approaches. New samples for systematic tests are constructed and described below. 

\subsection{Galaxy cluster catalog}
Our cluster member galaxy sample is taken from the SDSS DR8 \citep{Aihara11} redMaPPer v5.10 cluster
catalog\footnote{http://risa.stanford.edu/redmapper/}, constructed based on a red-sequence cluster
finding approach. Details of the algorithm and the cluster properties can be found in
\citet{Rykoff14, Rozo14,Rozo15a,Rozo15b}. Some features of the redMaPPer cluster catalog are briefly
summarized here. 

For each cluster, it identifies five most probable central galaxies, with their corresponding
central probability, $P\rm_{cen}$. Each cluster member galaxy is assigned with a membership
probability, $p\rm_{mem}$, according to its color, magnitude, and position information. The
photometric redshift $z$ for each cluster is estimated from high-probability members. The cluster
sample is approximately volume-limited in the redshift range of $0.1 \leq z \leq 0.35$. The cluster
richness,  $\lambda$, is defined by summing the membership probabilities over all possible cluster
members. Most of the clusters have $\lambda \gtrsim 20$, corresponding to an approximate halo mass threshold of $M\rm_{200m} \gtrsim 10^{14}\ h^{-1} M_{\odot}$ \citep{Rykoff12, Simet17}. 

\subsection{Galaxy shapes}

\subsubsection{Shape-related parameters}
\label{subsec:shape_par}

We adopt the following definition of ellipticity/distortion components in a global Cartesian frame to measure each galaxy's ellipticity:
\begin{equation}\label{eq:e1e2}
(e_1,\ e_2)=\frac{1-(b/a)^2}{1+(b/a)^2}(\rm{cos}\ 2\alpha,\ \rm{sin}\ 2\alpha),
\end{equation}
where $b/a$ is the minor-to-major axis ratio and $\alpha$ the position angle (PA) of the major axis of the galaxy. 
Here $e_1$ measures the projected distortion in the RA$/$dec directions, and $e_2$ in diagonal directions. 
The total galaxy ellipticity $e$ can then be calculated as
\begin{equation}\label{eq:e}
e=\sqrt{e_1^2+e_2^2}=\frac{1-(b/a)^2}{1+(b/a)^2}.
\end{equation}

\begin{figure}
\begin{center}
\includegraphics[width=0.42\textwidth]{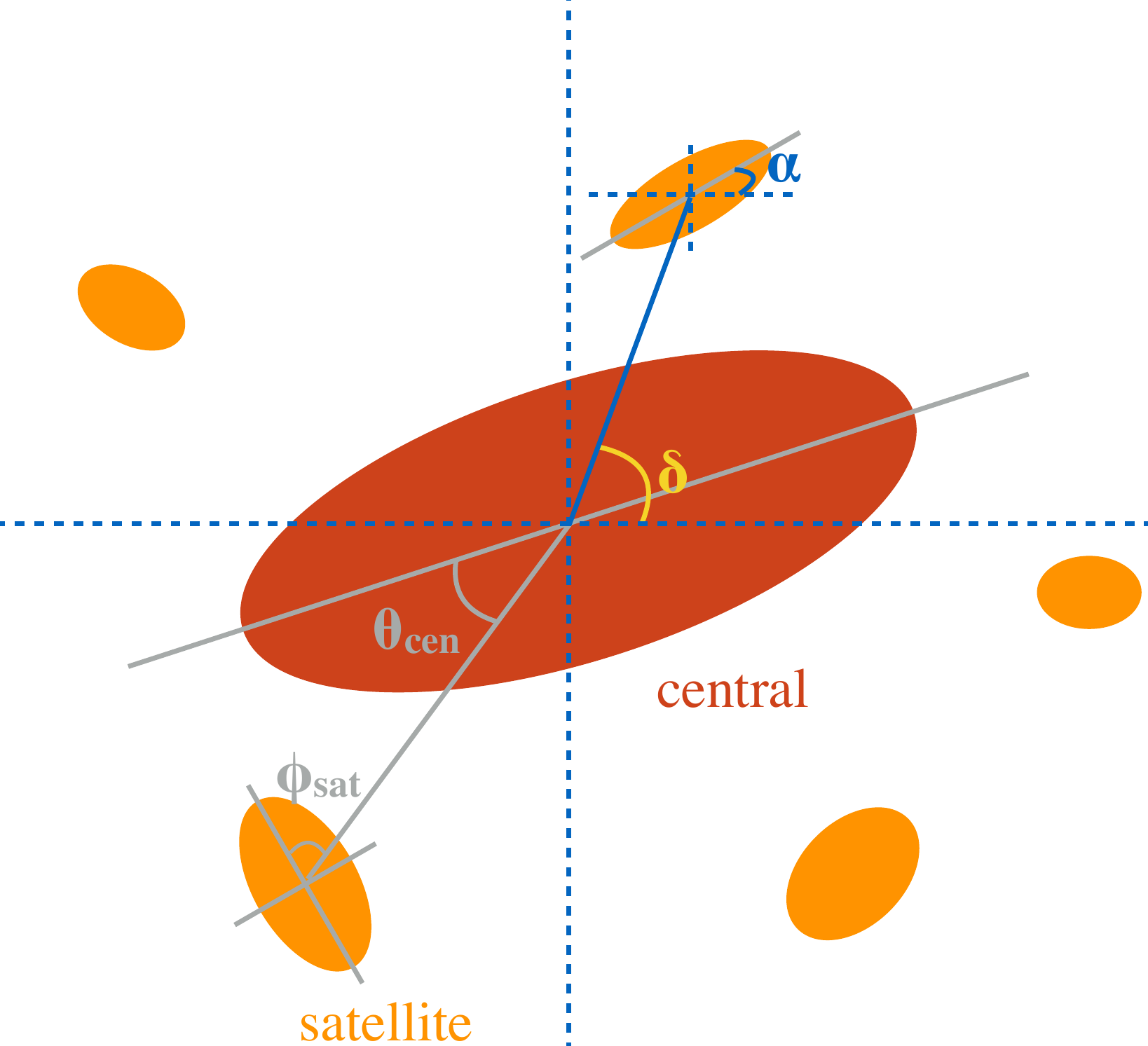}
\caption{Illustration of the galaxy alignment angles. The satellite alignment angle, $\phi_{\rm sat}$, defined as the angle between a satellite's major axis and its orientation towards the central galaxy, is the main focus of this paper. The central galaxy alignment angle, $\theta_{\rm cen}$, defined as the angular location of a satellite relative to its central galaxy's major axis direction, is the area of focus in our previous Paper I.}
\label{fig:illustration}
\end{center}
\end{figure}

Once the PA $\alpha$ of a galaxy is obtained by one of the methods described in Sec.~\ref{subsubsec:shapedata}, we can then derive its central galaxy alignment angle $\theta_{\rm cen}$ and satellite alignment angle $\phi_{\rm sat}$ as illustrated in Fig.~\ref{fig:illustration}. 

The central galaxy alignment angle ($\theta_{\rm cen}$) is defined as the angle between the major axis of the central galaxy and the line connecting the central to the satellite galaxy. 
We only need a viable shape measurement for the central galaxy (but not the satellites) to derive $\theta_{\rm cen}$. 
The analysis of central galaxy alignments in redMaPPer clusters has already been reported in Paper
I. 
The satellite alignment angle ($\phi_{\rm sat}$) is defined as the angle between the major axis of
the satellite galaxy and the line connecting its center to the central galaxy. Deriving $\phi_{\rm
  sat}$ requires a shape 
measurement for the satellite galaxy. In this paper, we focus on the satellite alignments, and will
use the central galaxy alignment angle $\theta_{\rm cen}$ as one of the candidate predictors in our
parameter pool. We use the highest-probability centrals provided in redMaPPer for calculation of both $\theta_{\rm cen}$ and $\phi_{\rm sat}$.

We restrict both $\theta\rm_{cen}$ and $\phi\rm_{sat}$ to the range [0$^\circ$,
90$^\circ$] due to symmetry. By definition, $\theta\rm_{cen}=0^\circ/90^\circ$ indicates a satellite
located 
along the major/minor axis of the central. A satellite is radially/tangentially aligned with the central if
$\phi\rm_{sat}=0^\circ/90^\circ$.

Besides using $\phi_{\rm sat}$ to quantify the degree of SA signal, another commonly-used parameter is $e_{\rm +}$, 
the distortion in the radial-tangential direction in a new frame with the original axes rotated to
the radial-tangential directions of each central-satellite pair. 
From simple algebra, we have:
\begin{equation}\label{eq:eret}
(e_{\rm +},\ e_{\rm \times})=\frac{1-(b/a)^2}{1+(b/a)^2}(\rm{cos}\ 2(\delta-\alpha),\ \rm{sin}\ 2(\delta-\alpha)),
\end{equation}
where $\alpha$ is the PA of the satellite, and $\delta$ the azimuthal angle of the satellite projected position with respect to the cluster central galaxy, as indicated in Fig.~\ref{fig:illustration}.
A positive $e_{\rm +}$ indicates a radial alignment of the satellite toward cluster center, while a
negative $e_{\rm +}$ indicates a tangential alignment. Therefore, if satellite galaxies do
preferentially align in the radial direction, we expect $\mean{e_{\rm +}}>0$ when taking the average over all central-satellite pairs. 
The $e_{\rm \times}$ component is the distortion at $\pm 45^\circ$ from the radial/tangential
direction. It is is commonly used as an indicator for certain systematics. Due to symmetry, $\mean{e_{\rm \times}}$ should be consistent with zero.

\subsubsection{Shape data}\label{subsubsec:shapedata}

We will measure satellite alignments using three shape measurement methods: re-Gaussianization,
isophotal, and de Vaucouleurs shapes, to compare differences in the signals and investigate
systemics. Details of these methods have been described in Paper I (Sec.~2.3), and we only briefly
summarize here.

The re-Gaussianization shape measurement method \citep{Hirata03} is specifically designed for weak
lensing studies, which require great care in removing the point spread function (PSF) effect on the
observed galaxy images. This method has a Gaussian weight function that emphasizes the inner, brighter regions of galaxy profiles in order to 
reach higher precision distortion measurement especially for faint galaxies. 
In this work, we take the distortion measurement ($e_1$ and $e_2$) from a re-Gaussianization shape
catalog of \citet{Reyes12} with shapes measured in the $r$ and $i$ bands based on the SDSS DR8 photometry. 

Isophotal shape measurement does not include an explicit correction for the effect of the PSF. It determines a galaxy's shape by fitting the surface brightness at 25 mag/arcsec$^2$, which traces the outer part of a galaxy's profile. Since isophotal shapes were not released in DR8, we take the isophotal position angle in $r$ band from DR7 to compute satellite alignment angles. 

The de Vaucouleurs shape measurements were determined by fitting each galaxy's image with a de
Vaucouleurs model \citep{Stoughton02}, which is a good description for typical elliptical galaxies
(which includes the majority of the galaxies in this work, since they were selected based on a
red-sequence method). It partially corrects for the PSF effect using an approximate PSF model, and
overall is sensitive to light profiles on scales between those measured by re-Gaussianization and
isophotal methods. 
We use the de Vaucouleurs fit position angle in $r$ band provided from the SDSS DR7 in this work
\footnote{We have tried fixing the shape measurement method to re-Gaussianzation but varying SDSS photometry pipeline between DR4 and DR8. We found that the derived shape parameters based on different pipelines are statistically consistent. Similarly, we expect that using DR7 photometry for de Vaucouleurs shapes should give consistent results as using those based on DR8. 
}. 

\subsection{The central-satellite pair sample}
\label{subsec:sample1}

To fairly compare the measured alignment signal across redshift, we restrict our analysis to a volume-limited cluster sample within $0.1 \leq z \leq 0.35$ from the redMaPPer catalog. 
Besides this, an appropriate membership probability cut of $p\rm_{mem} \geq 0.55$ is applied on
satellite galaxies, which results in a total of 305997 central-satellite pairs in 10749 distinct
clusters (before requiring galaxies to have shape measurements). 
The choice of the $p\rm_{mem} = 0.55$ cut comes from optimizing the $S/N$ for detection of
SA signals, as explained in Appendix~\ref{app:Pmem cut}.

While applying a lower $p_{\rm mem}$ cut returns us more central-satellite pairs into analysis, the
resulting satellite alignment signal will be diluted due to the inclusion of more pairs with
``satellite'' galaxies that are not actually in clusters. 
Throughout this work, we will reduce this contamination by applying $p_{\rm mem}$ as the
weighting factor on each central-satellite pair.

We define two sets of central-satellite pair samples for analysis in this work.
\begin{enumerate}

\item \textbf{DR8 footprint sample}: The first set is within the SDSS DR8 footprint, constructed by acquiring that the 305997 $p\rm_{mem} \geq 0.55$ satellites have well-defined re-Gaussianization shape measurements. 
There are 174180 central-satellite pairs within 8121 distinct clusters in this data set. 
The effective total number of pairs in DR8 footprint sample after weighting by $p_{\rm mem}$ is $\sum\limits_{i} p_{{\rm mem},i} \approx 132072$ pairs. 

\item \textbf{DR7 footprint sample}: The second data set is constructed for comparing the level of satellite alignment signals via three
different shape measurement methods. We require satellites in this subsample to have all three kinds
of shape measurements, and thus this data set covers the smaller DR7 footprint. In total, there are
158537 central-satellite pairs within 7385 distinct clusters, or effectively 120200 after weighting
by $p_{\rm mem}$.
\end{enumerate}

The resulting redshift and luminosity distributions of satellites in the constructed DR8 and DR7 sample sets are almost indistinguishable,
indicating that the selection functions for different shape measurements are quite similar. 

In Paper I, to investigate the effect of the sky-subtraction technique on the measured central
galaxy alignment signal, we have reported the results based on a set of DR4 footprint satellites,
which have re-Gaussianianization shape measurement based on both DR4 and DR8 SDSS photometric
pipelines. We found that within error bars, the measured central galaxy alignment signals are very
similar when different photometric pipelines are used, and therefore concluded that the effect of
sky-subtraction does not substantially influence the central galaxy alignment measurement. In this
work for satellite alignment, we have also examined whether sky-subtraction is an issue for the
detecting signal using the DR4 footprint data set, but did again failed to find any
disagreement. For this reason, to simplify the analysis we will only report the measured satellite alignment results for the DR8 and DR7 footprint sample.

\subsection{Systematic test sample}
\label{subsec:sample2}

We construct three systematic test samples to study potential systematic effects in the crowded cluster environment. 
All of the samples are constructed from the red-sequence Matched filter Galaxy v6.3 Catalog
(redMaGiC) \citep{Rozo16}, a photometrically-selected luminous red galaxy (LRG) catalog with very
high quality photo-$z$ estimation based on the SDSS DR8 photometric data. 
Overall, the bias, defined as the median value of $z_{\rm photo}-z_{\rm spect}$, of the DR8 redMaGiC photo-$z$ is less than 0.005. 
The $1\sigma$ scatter of $(z_{\rm photo}-z_{\rm spect})/(1+z_{\rm spect})$ is $\lesssim$ 0.02.

\begin{enumerate}

\item \textbf{Foreground \& background of redMaPPer}: The foreground and background sample is
  composed of galaxies that are in the same sky area as redMaPPer clusters, but are not physically
  associated with the cluster. This sample is constructed as follows. For each central galaxy in the redMaPPer cluster, we use 1.5$R_c(\lambda)$ as a searching radius to select out LRGs within the projected area in the redMaGiC catalog. 
The $R_c(\lambda)$ is the radius within which $p_{\rm mem}$ is assigned in the original redMaPPer catalog, and it is estimated that $R_{\rm 200c}\approx1.5R_c(\lambda)$ \citep{Rykoff12,Rykoff14}.
Next we select LRGs whose photo-$z$ $<$ ($>$) $z_{\rm cluster}$ as foreground (background)
candidates.
To improve the purity of the sample, 
we further exclude $p_{\rm mem} > 0.2$ galaxies in the ``ubermem''
version of the redMaPPer catalog.  This ``ubermem'' catalog extends the $p_{\rm mem}$ estimation out to 
$1.5R_c(\lambda)$ and down to fainter galaxies ($mag\ i < 21$), thus enabling us to 
remove potential cluster members at $1\sim1.5R_c(\lambda)$ and those that are relatively faint,
unlike in the original redMaPPer catalog. 
After this cut, 95\% of foregrounds and 99\% of backgrounds have their $\Delta z = |z_{\rm
  photo}-z_{\rm cluster}| \gtrsim 0.02$, suggesting that the above procedures return a set of clean
foreground and background galaxies. 
After requiring these galaxies to have re-Gaussianization shape measurements, we have 45030 fake
central-satellite pairs in the DR8 footprint, of which 4459 and 40571 are foreground and background
galaxies, respectively. 
Further requiring these galaxies to have de Vaucouleurs and isophotal shapes in DR7, leaves us with
4134 and 36941 foreground and background galaxies, respectively.

\item \textbf{Non-cluster field sample}: To highlight the effect of the crowded cluster environment
  on the shape measurement of galaxies within the cluster area on the sky (either
  physically-associated member galaxies or just foreground/background galaxies), we construct a
  sample of galaxies that are not in the footprint of redMaPPer cluster fields, and call it the
  ``non-cluster field sample''.  We do so by simply taking the full redMaGiC catalog, and excluding
  all galaxies ($p_{\rm mem} \geq$ 0) that belong to the redMaPPer cluster member sample 
  as well as galaxies in the foreground and background sample constructed above. 
    After requiring that these galaxies
  have all three shape measurements, we have 697308 galaxies within the DR7 footprint. 

\item \textbf{Foreground \& background of $m_r$<19 non-cluster field bright galaxies}: 
In order to understand the level of contamination caused by the extended light profile of bright
central galaxies on nearby satellites, we select  bright galaxies with $m_r$<19 from our non-cluster
field sample described above, and construct a sample composed of the corresponding foreground and background galaxies around these bright galaxies. 
We again find the foreground/background galaxies from redMaGiC, and require their $z_{\rm photo}$ to
be at least $0.04$ smaller/greater than that of their nearby bright galaxies. Here 0.04 is chosen to
be about $2\sigma$ given the average photo-$z$ error of redMaGiC. 
Requiring that all foreground and background galaxies having well-defined shape measurements yields 281114 galaxies in total. 

\end{enumerate}

Table~\ref{tb:samples} summarizes the data sets we have defined in Secs.~\ref{subsec:sample1} and~\ref{subsec:sample2}, with detailed sample size information provided.

\begin{table*}
\caption{Sample sets used in this work. 
The upper part of the table shows the two cluster subsamples used for the overall measurement of the satellite alignment effect, as described in Sec.~\ref{subsec:sample1}. 
The columns indicate the total number of satellites (N$_{\rm sat}$), total number of distinct
clusters (N$_{\rm cluster}$), and the effective number of central-satellite pairs for each sample
after weighting by $p_{\rm mem}$ (N$_{\rm eff}$).
The lower part of the table summarizes the subsamples used for systematic tests, as described in Sec.~\ref{subsec:sample2}. 
The columns indicate the total number of galaxies (N$_{\rm tot}$), number of foreground (N$_{\rm fore}$) and background objects (N$_{\rm back}$), respectively. 
}
\begin{tabular}{lllll}
\hline 

\textbf{Cluster system sample}				&	N$_{\rm sat}$ 		&   N$_{\rm cluster}$ 	&	N$_{\rm eff}$    \\  \hline  

DR8 footprint sample					&	174180			&	8121				&	132072		\\	
DR7 footprint sample					&	158537			&	7385				&	120200		\\  \hline

\\ \hline
\textbf{Systematic test sample}				& 	N$_{\rm tot}$		& N$_{\rm fore}$		& N$_{\rm back}$  	\\ \hline

Foreground \& background of redMaPPer	(DR8 footprint)	&	45030 			&  4459  				& 40571 			\\ 

Foreground \& background of redMaPPer	(DR7 footprint)	&	41075 			&  4134  				& 36941 			\\ 
	
Non-cluster field sample							&  	697308 			&  			   						\\

Foreground \& background of $m_r$<19 non-cluster field bright galaxies 	& 278204 & 6990		& 271214					\\
\hline

\label{tb:samples}
\end{tabular}
\end{table*}

\subsection{Summary of physical parameters}
\label{subsec:parameters}

Similar to our analysis in Paper I, we aim to identify predictors that significantly influence the satellite alignment effect from a large parameter pool. 
Almost all of the physical parameters explored in this work are the same as in Paper I, except for one newly added variable: \textit{fracDeV}.

The \textit{cmodel} magnitude systems in SDSS pipeline tries to fit galaxy light profile by taking the linear combination of both de Vaucouleurs and exponential profiles, 
and stores the coefficient of the de Vaucouleurs term in the quantity \textit{fracDeV}, which describes the fraction of light from a fit to a de Vaucouleurs profile. 
For a galaxy that can be best fitted by pure exponential profile, \textit{fracDeV} = 0, while \textit{fracDeV} = 1 for pure de Vaucouleurs profile. 
In general, the brightness distribution of disks follows the exponential profile, whereas bulges are
better described with a de Vaucouleurs profile. The \textit{fracDeV} parameter thus can be viewed as
a tracer for a galaxy's angular momentum content or its overall morphology, which has similar but
not identical information to galaxy color. It is interesting to check the dependence of angular momentum on SA signal.

In total, we have one response variable ($\phi_{\rm sat}$) and 17 other variables constituting the pool
of possible predictors for $\phi_{\rm sat}$.
We classify the 17 parameters into three categories: satellite-related quantities, central galaxy-related quantities and cluster-related quantities.
A brief summary of important information about these parameters is in Table~\ref{tb:predictors}; we refer readers to Sec.~2.2 of Paper I for details. 

\begin{table*}
\caption{A summary of the 17 parameters used to study the satellite alignment effect.}
\begin{center}
\begin{tabular}{lc}
\hline 

\textbf{Response Variable}  	& \textbf{Properties} \\ \hline

$\phi_{\rm sat}$					& \makecell[l]{ $\bullet$ Satellite alignment angle, as demonstrated in Fig.~\ref{fig:illustration}. 					\\
									       $\bullet$ We use this parameter as a response variable to quantify the level of satellite alignment.	\\
									       	Smaller $\phi_{\rm sat}$ indicates a stronger satellite alignment effect. } 	\\
\hline

\\
\\ \hline

\textbf{Satellite Galaxy Quantities} 	& \textbf{Properties} \\ \hline
log($r/R_{\rm 200m}$)			& \makecell[l]{$\bullet$ Member distance from the cluster central galaxy, normalized by $R_{\rm 200m}$} 			\\ \\
satellite $^{0.1}M_r$				& \makecell[l]{$\bullet$ r-band absolute magnitude of the satellite, k-corrected to $z=0.1$}						\\ \\
satellite $^{0.1}M_g-^{0.1}M_r$ 	& \makecell[l]{$\bullet$ Color of the satellite galaxy, k-corrected to $z=0.1$}													\\ \\
satellite ellipticity 				& \makecell[l]{$\bullet$ Satellite ellipticity as defined in Eq.~\ref{eq:e}} 									\\ \\
$\Delta$log(satellite $R_{\rm eff}$)	& \makecell[l]{ $\bullet$ The excess of galaxy size with respect to the predicted size at the same luminosity 		\\
										$\Delta$log(satellite $R\rm_{eff}$) $\equiv$ measured log(satellite $R\rm_{eff}$) $-$ predicted log(satellite $R\rm_{eff}$) \\
									       $\bullet$ The predicted log(satellite $R\rm_{eff}$)$= -0.20$(satellite $^{0.1}M_r$)$-3.84$, derived by linearly fit to all  \\
									       $p_{\rm mem}>0.55$ satellites in the DR8 footprint sample. A relevant figure is presented in Fig.~5 of \\
									       Paper I, except that the derived predicted log(satellite $R\rm_{eff}$) is slightly different from Paper I due \\
									       to the inclusion of lower $p_{\rm mem}$ satellites in Paper II.} 					\\ \\
\textit{fracDeV}					& \makecell[l]{ $\bullet$ The fractional flux contribution of the de Vaucouleurs profile, see Sec.~\ref{subsec:parameters} for detail. \\
									       $\bullet$ \textit{fracDeV}=0 for a pure exponential profile; \textit{fracDeV}=1 for a pure de Vaucouleurs profile.
									       }      \\ \\									      
$\theta_{\rm cen}$				& \makecell[l]{ $\bullet$ Central galaxy alignment angle, as demonstrated in Fig.~\ref{fig:illustration} 				\\
									       $\bullet$ Smaller $\theta_{\rm cen}$ indicates that the
                                           satellite is residing closer to the major axis direction of its \\
										central galaxy.} \\ 

\hline

\\
\\ \hline

\textbf{Central Galaxy Quantities} 	& \textbf{Properties} \\ \hline

central galaxy dominance			& \makecell[l]{$\bullet$ Magnitude gap between the central galaxy and the mean of the 2nd and 3rd brightest satellites	\\
								${\rm Central\ dominance}\  \equiv\ {\rm Central}\ ^{0.1}M_{\rm r} - \frac{^{0.1}M_{\rm r,1st}+^{0.1}M_{\rm r, 2nd}}{2}$ \\
									      $\bullet$ A smaller value indicates a more dominant central galaxy.}					\\ \\

central $^{0.1}M_r$ 				& \makecell[l]{$\bullet$ r-band absolute magnitude of the central, k-corrected to $z=0.1$} 						\\ \\

central $^{0.1}M_g-^{0.1}M_r$		& \makecell[l]{$\bullet$ Color of the central galaxy, k-corrected to $z=0.1$} 				      									\\ \\

central ellipticity				&  \makecell[l]{$\bullet$ Ellipticity of central galaxy as defined in Eq.~\ref{eq:e}.} 				      				\\ \\

$\Delta$log(central $R_{\rm eff}$)	& \makecell[l]{ $\bullet$ The excess of central galaxy size with respect to the predicted size of centrals 			\\
											    at the same luminosity. 														\\
										$\Delta$log(central $R\rm_{eff}$) $\equiv$ measured log(cental $R\rm_{eff}$) $-$ predicted log(central $R\rm_{eff}$) \\
									       $\bullet$ The predicted log(central $R\rm_{eff}$)$= -0.31$(central $^{0.1}M_r$)$-6.16$, derived by linearly fit to all \\
									       centrals in the DR8 footprint sample. See also Fig.~4 in Paper I for more more detail. } 			\\ \\

$P_{\rm cen}$					& \makecell[l]{ $\bullet$ Central galaxy probability provided in redMaPPer. \\										 									       $\bullet$ $P_{\rm cen}$ is an indicator of whether a cluster system contains only a single dominant central \\
									       galaxy or has multiple central galaxy candidates.} 								     \\ 
\hline

\\
\\ \hline
\textbf{Cluster Quantities} 			& \textbf{Properties} \\ \hline	
log(richness)					& \makecell[l]{$\bullet$ Cluster richness taken from the redMaPPer catalog.}									\\ \\
redshift						& \makecell[l]{$\bullet$ Cluster redshift estimated by redMaPPer.}											\\ \\
cluster ellipticity					& \makecell[l]{$\bullet$ Calculated based on the distribution of member galaxies. See Sec. 2.2.3 of Paper I for detail.		\\
									      $\bullet$ A cluster with larger cluster ellipticity has more elongated satellite distribution.}				\\ \\
cluster member concentration, $\Delta_{\rm R}$	& \makecell[l]{$\bullet$ Derived based on the average projected distance of member galaxies from the cluster center, \\
												     with some normalization towards cluster richness and redshift. See Sec. 2.2.9 of Paper I for  	\\
												     detail definition.	\\
												     $\bullet$ By construction, negative $\Delta_{\rm R}$ value means the cluster has a more compact member galaxy\\
												      distribution than the average cluster at similar richness and redshift.}	\\
\hline

\label{tb:predictors}
\end{tabular}
\end{center}
\end{table*}


%% file: 3_SAsignal.tex
\section{Overall Signal of Satellite Alignment}
\label{sec:sat}


\subsection{Distribution of $\phi_{\rm sat}$}
\label{subsec:DR8_phi_sat}

\begin{figure}
\begin{center}
\includegraphics[width=0.49\textwidth]{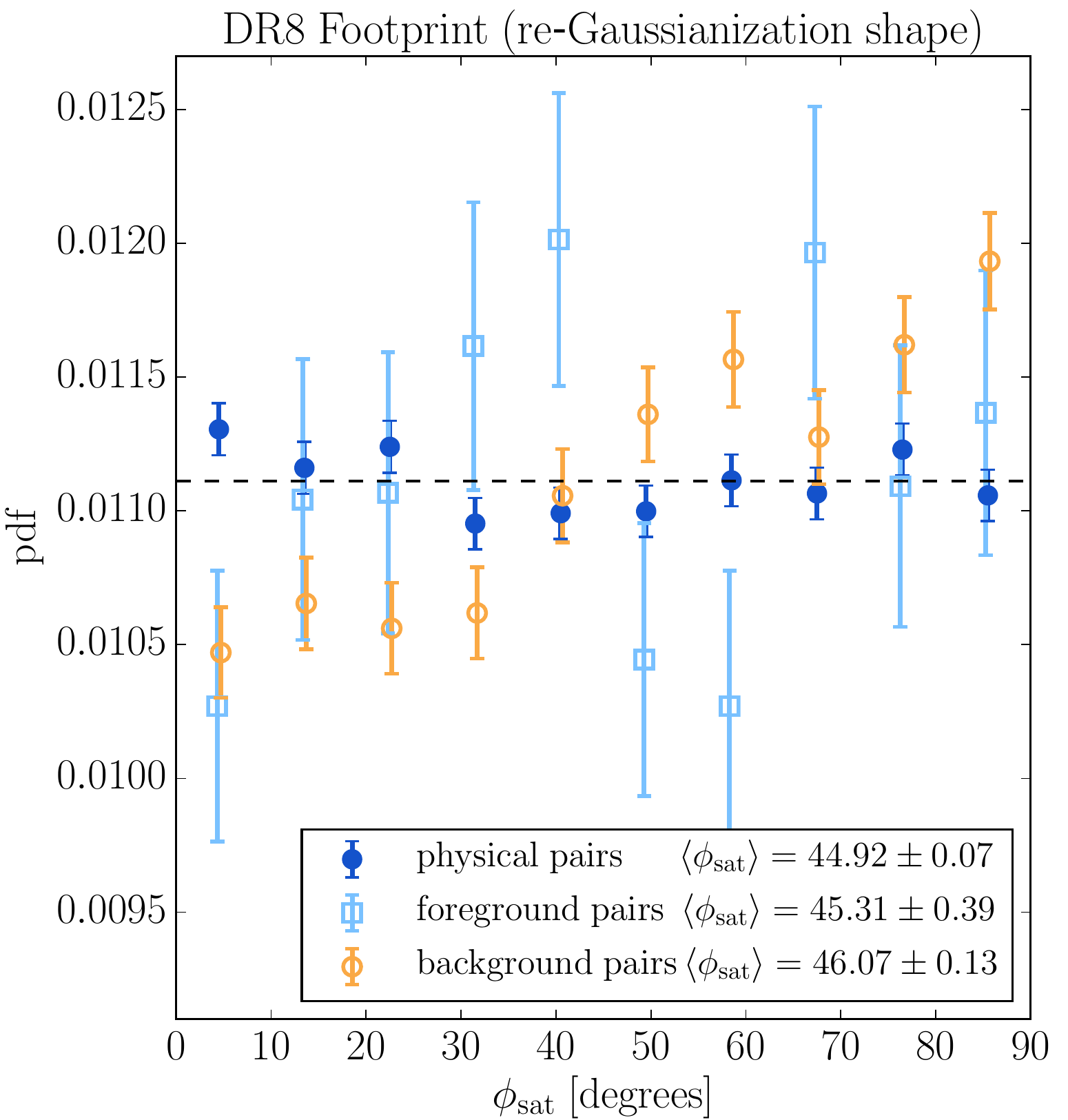}
\caption{$p_{\rm mem}$-weighted probability density distribution of SA angle, $\phi_{\rm
    sat}$, for the 174180 $p_{\rm mem} > 0.55$ pairs measured using the re-Gaussianization method. The dark
  blue filled circles indicate the pdf of $\phi_{\rm sat}$ for central-satellite pairs of redMaPPer clusters
  selected using $p_{\rm mem} > 0.55$. 
  \textit{The weighted averaged SA angle, $\mean{\phi_{\rm sat}}$, is consistent with 45$^\circ$ within error bar, indicating that the overall SA signal measured based on re-Gaussianization shape is not very significant.}
  The open markers show the probability density function (pdf) of $\phi_{\rm sat}$ for
  foreground (light blue square) and background (orange circle) pairs in the footprint of redMaPPer clusters. 
  As expected, the $\mean{\phi_{\rm sat}}$ value for foreground is consistent with random, and there is tangential alignment signal $\mean{\phi_{\rm sat}} > 45^\circ$ for backgrounds.
  The horizontal black dashed line shows the expected result for randomly oriented satellite galaxies. 
} 
\label{fig:phisat_dist_DR8}
\end{center}
\end{figure}

The dark blue filled circles in the left panel of Fig.~\ref{fig:phisat_dist_DR8} indicate the $p_{\rm
  mem}$-weighted distribution of the SA angle, $\phi_{\rm sat}$, for our 174180 satellites in the
DR8 footprint, based on the re-Gaussianization shape measurements. The weighted average SA angle is
$\mean{\phi_{\rm sat}} = 44.92^\circ \pm 0.07^\circ$, consistent with no net tendency for SA.

However, it does not rule out the possibility
of a statistically significant SA detection for subsamples of the satellite population. We will show
later in Sec.~\ref{subsec:LR-reG} that when focusing on brighter satellites, we can still detect a
statistically significant SA signal using re-Gaussianization shapes.

\subsection{SA measurement in $e_{\rm +}$}
\label{subsec:e+}

Besides using the parameter $\phi_{\rm sat}$ to quantify the degree of SA signal, it is also useful
to calculate the mean radial ellipticity $\mean{e_{\rm +}}$, especially when quantifying the level
of IA systematics to weak lensing signals \citep[e.g.,][]{Schneider13, Sifon15}. Here we also
compute mean radial ellipticities in order to compare with previous work. 
For the definition of $e_{\rm +}$, we refer readers back to Sec.~\ref{subsec:shape_par} for more
detail. Under our definition, a satellite with $e_{\rm +} > 0$ tends to point radially toward its host central galaxy. 

\begin{figure}
\begin{center}
\includegraphics[width=0.48\textwidth]{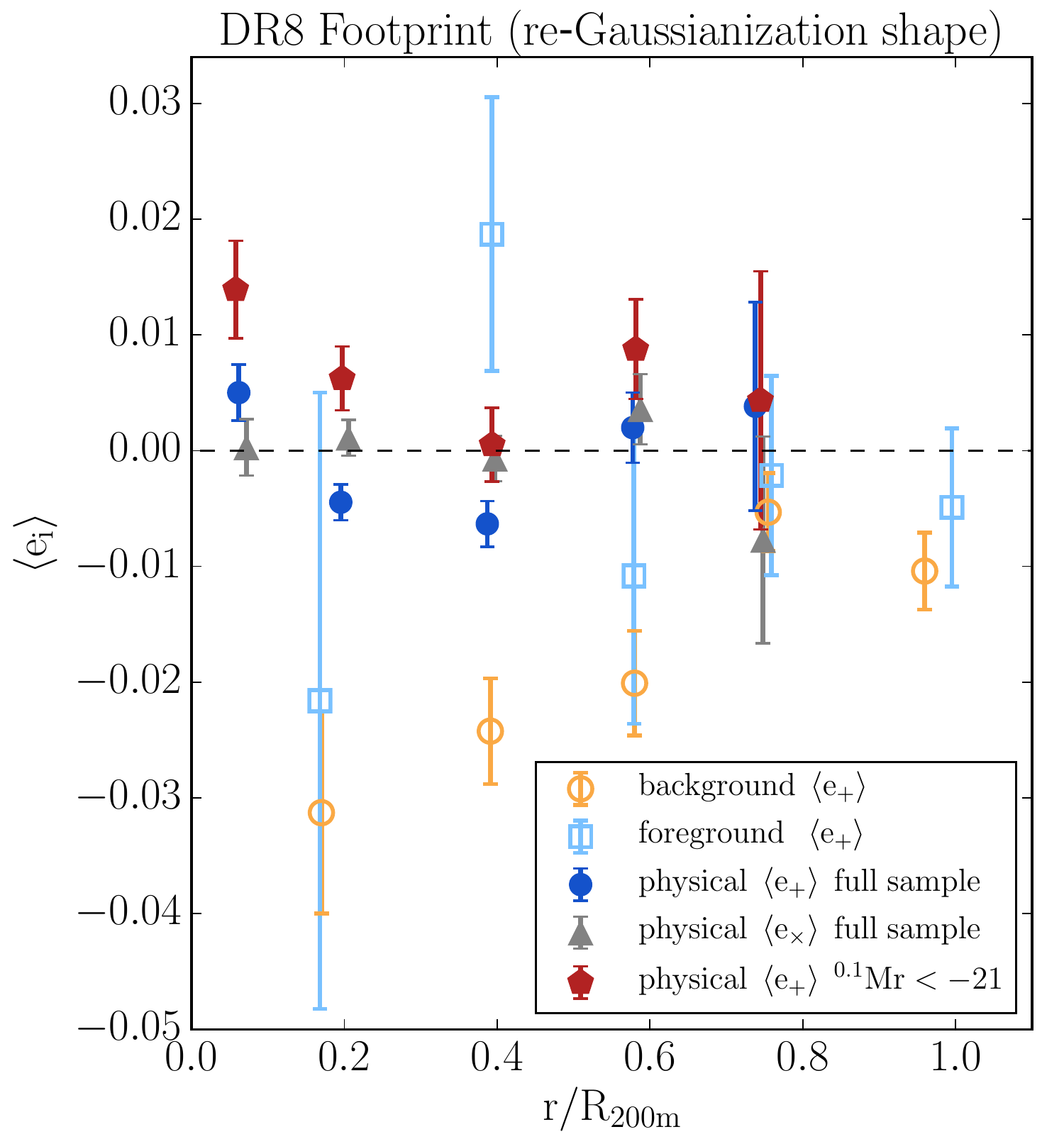}
\caption{Satellite alignment signal measured in $\mean{e_{\rm +}}$ (dark blue filled circle) in bins of normalized
  projected distance based on re-Gaussianization shapes for the 174180 DR8 redMaPPer $p_{\rm mem}>0.55$ satellites. 
  Under our definition, $\mean{e_{\rm +}} > 0$ indicates radial alignment.
  \textit{The measured SA signal is consistent with zero within 3$\sigma$ across all radial bins.}
The $\mean{e_{\rm +}}$ signals for foregrounds (light blue square) and backgrounds (orange open circle) in redMaPPer cluster fields are also shown. 
The $\mean{e_{\rm \times}}$ component for physical pairs, indicated in grey triangles, is consistent
with zero in all radial bins. \textit{This suggests that systematics that would cause a B-mode
  signal  in the re-Gaussianization shapes are negligible.}
When focusing on a subsample of brighter satellites with $^{0.1}Mr < 21$ (red pentagon), we find
$\mean{e_{\rm +}}\approx0.014\pm0.0042$ ($\sim$3.3$\sigma$) and $\mean{e_{\rm
    +}}\approx0.0062\pm0.0028$ ($\sim$2.3$\sigma$) in the two smallest radial bins at $\sim$ $r <
0.2 R_{\rm 200m}$. \textit{This indicates that we reach a significant SA detection using re-Gaussianization shapes with satellites that are more luminous and located closer to central galaxies.}
}
\label{fig:er_dist}
\end{center}
\end{figure}

The dark blue filled circles in Fig.~\ref{fig:er_dist} show the averaged $\mean{e_{\rm +}}$ component based on
re-Gaussianization shapes for all DR8 footprint satellites, divided into bins in projected
separation from their own central galaxy. The corresponding error bars are simply the standard error of the mean. 
We observe that the SA signal is consistent with zero
within 3$\sigma$ across all radial bins, meaning that we do not detect any significant SA effect in
the overall redMaPPer satellite population.  
The grey triangles indicate the measured averaged ellipticity component $\mean{e_{\rm \times}}$, which provides a 45$^\circ$ systematic test (also known as B-mode test). 
By symmetry, the expected $\mean{e_{\rm \times}}$ value should be zero, unaffected by either lensing or SA effects, which only contribute to the $e_{\rm +}$ component.
We find that our measured $\mean{e_{\rm \times}}$ is consistent with zero in all radial bins,
suggesting that systematic errors that would generate a B-mode signal are negligible.
  
While there is no coherent radial orientation of satellites across the entire DR8 sample, as we will show later in Sec.~\ref{subsec:LR-reG}, more luminous satellites tend to have stronger SA signal. 
Here we demonstrate that a subsample of satellites with $^{0.1}Mr < -21$ has $\mean{e_{\rm +}}\approx0.014\pm0.0042$ ($\sim$3.3$\sigma$) and $\mean{e_{\rm +}}\approx0.0062\pm0.0028$ ($\sim$2.3$\sigma$) in the two smallest radial bins at $\sim$ $r < 0.2 R_{\rm 200m}$, as shown in the red pentagons of Fig.~\ref{fig:er_dist}. 
In comparison, \citet{Sifon15} found no significant SA across all
radial bins for satellites with $^{0.1}Mr < -21$ (see their Fig.~10) when using satellites of 91 massive galaxy clusters with shape measurements optimized for lensing. 
Similarly, \citet{Schneider13} also found no apparent SA signal across all radial bins for
early-type satellites based on members of galaxy groups (see their Fig.~7).

\subsection{SA signal based on different shape measurements}
\label{subsec:DR7_phi_sat}

\begin{figure}
\begin{center}
\includegraphics[width=0.49\textwidth]{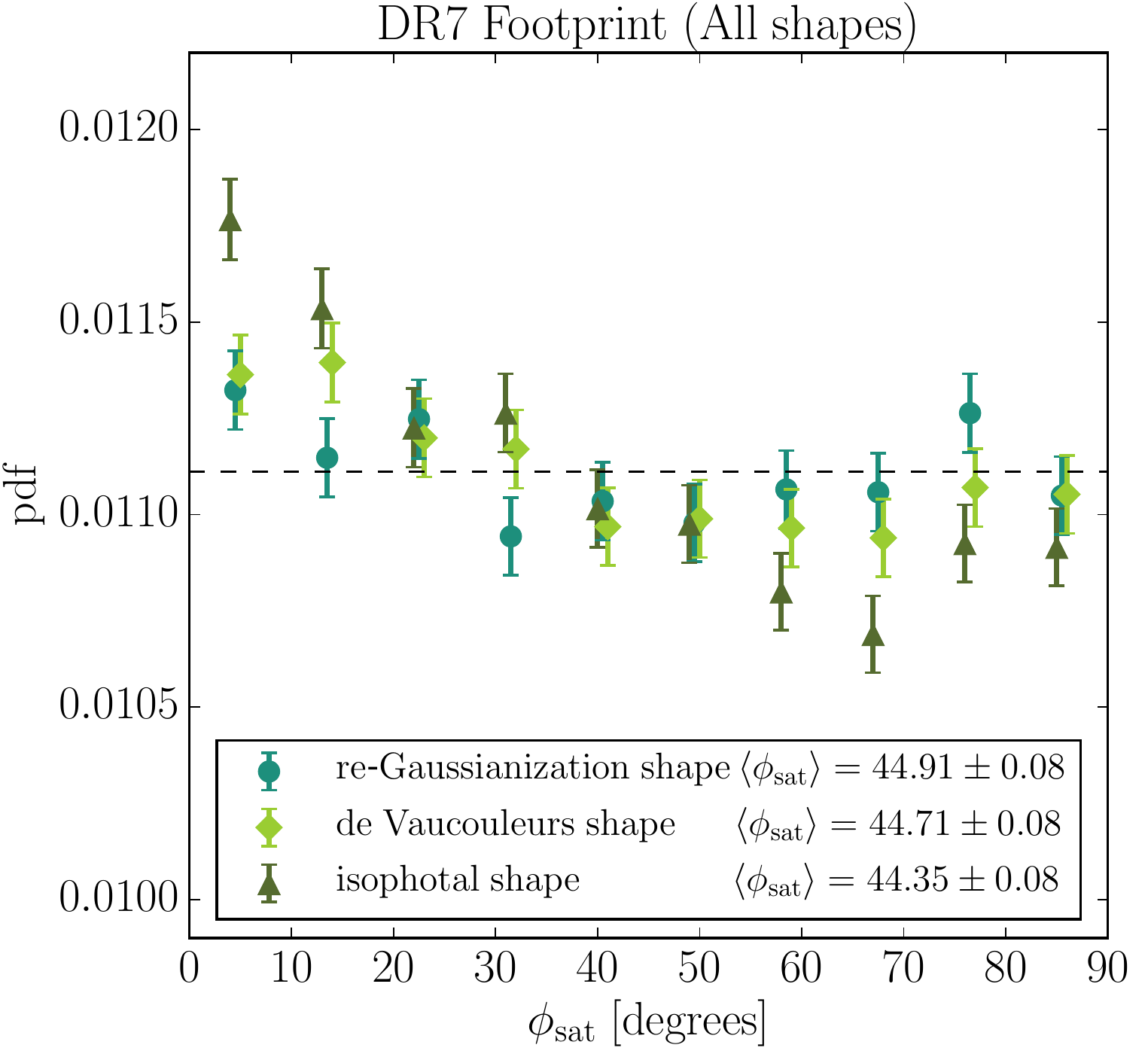}
\caption{$p_{\rm mem}$-weighted probability density distribution of $\phi_{\rm sat}$ for
  the 158537 $p_{\rm mem} > 0.55$ redMaPPer satellites in the DR7 footprint based on re-Gaussianization
  (teal green circle), de Vaucouleurs (yellow green diamond), and isophotal (olive triangle) shape measurements.
  \textit{The weighted average SA angles, $\mean{\phi_{\rm sat}}$, for both de Vaucouleurs and isophotal shapes
   are less than 45$^\circ$, indicating we have observed SA effect significantly based on these two shape measurements.
   However for re-Gaussianization shape, $\mean{\phi_{\rm sat}}$ is consistent with 45$^\circ$ within error bars.
   } }
\label{fig:phisat_dist_DR7}
\end{center}
\end{figure}

To investigate the effect of shape measurement methods on the detection of SA signals, we use
satellites within the DR7 footprint, as defined in Sec.~\ref{subsec:sample1} (see also
Table~\ref{tb:samples}), which have re-Gaussianization, deVaucouleurs, and isophotal shape
measurements. The left panel of Fig.~\ref{fig:phisat_dist_DR7} shows the $p_{\rm mem}$-weighted
distribution of $\phi_{\rm sat}$ for this sample set, with the teal green circles, yellow green diamonds and olive
triangles representing shape measurements based on the re-Gaussianization method, de Vaucouleurs fits,
and isophotal fits, respectively. The isophotal shape measurement produces the strongest SA signal
($\mean{\phi_{\rm sat}} =44.35^\circ \pm 0.08^\circ$), followed by de Vaucouleurs fits
($\mean{\phi_{\rm sat}} =44.71^\circ \pm 0.08^\circ$) and finally re-Gaussianization shapes ($\mean{\phi_{\rm sat}} = 44.91^\circ \pm 0.08^\circ$). 
However, as described in \citet{Hao11}, we still must test whether the detected SA signal is
  real (due to physical alignments) or fake (due to systematics). 
We will investigate possible systematic effects in Sec.~\ref{sec:cp_shape}.

\subsection{Foreground and background systematic tests}

We examine our SA measurement using sample sets of foreground and background galaxies in the footprint of redMaPPer cluster field. 
For construction of foreground and background samples, we refer readers back to Sec.~\ref{subsec:sample2}. 
For foregrounds, we expect galaxies to be randomly oriented in the measured $\phi_{\rm sat}$ with
respect to the central galaxies of redMaPPer clusters in the same field.
For backgrounds, we expect galaxies to exhibit tangential alignment because of the gravitational lensing effect. 
The light blue sqares/orange open circles in Fig.~\ref{fig:phisat_dist_DR8} show the distribution of
$\phi_{\rm sat}$ measured using re-Gaussianization shapes for our foreground/background samples. The
observed $\phi_{\rm sat}$ distributions are consistent with our expectation. 
This indicates that there are no severe systematics due to the complexity of
measuring shapes in cluster fields based on re-Gaussianization method. However, the test we applied is not very sensitive for low-level systematics due to the lack of foreground pairs. 

\begin{figure*}
\begin{center}
\includegraphics[trim=0.15cm 0cm 1.0cm 0cm,width=0.95\textwidth]{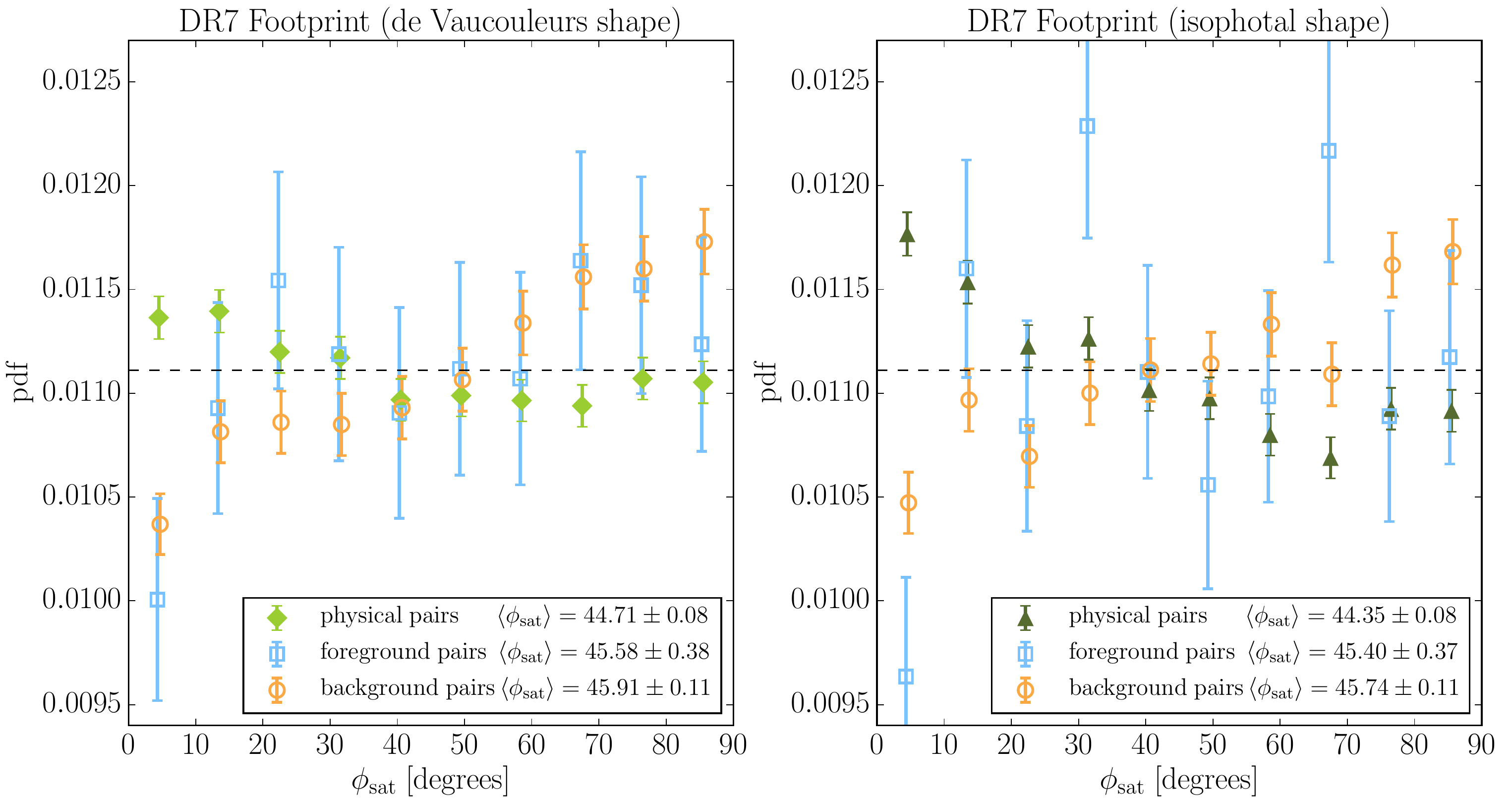}
\caption{Foreground (light blue open square) and background (orange open circle) tests for de
  Vaucouleurs (left panel) and isophotal shape (right panel) measurements.}
\label{fig:dr7fb_phisat_dist}
\end{center}
\end{figure*}

Besides the test for re-Gaussianization shape, Fig.~\ref{fig:dr7fb_phisat_dist} shows the foreground (light blue square) and background (orange open circle) tests for de Vaucouleurs shape (left panel) and isophotal shape (right panel). 
For foregrounds, the $p$-values of KS tests, as indicated in the legend below the figures, show that the distribution is consistent with uniform distribution. For backgrounds, we also observe the expected lensing effect in de Vaucouleurs and isophotal shaps. 

%% file: 4_LR.tex
\section{Linear Regression Analysis}
\label{sec:LR}

We apply linear regression analysis and variable selection techniques to properly account for correlations among various parameters and to identify featured predictors that significantly affect the SA phenomenon. 
The variable selection methods are quite similar (but not identical) to those described in Sec.~3 of Paper I. 
Below, we briefly summarize the approaches, including the new methodology in this paper, and report the results. 

\subsection{Methodology}
\label{subsec:method}

\subsubsection{Overview of Linear Regression}

Linear regression is a method to study the relationship between a response variable $Y$ and a variety of regressors vectorized as $\bold{X}= (X_1, X_2, X_3, ... ,X_N)$. One tries to estimate optimal values of the free parameters by minimizing the squared residuals of the following model: 
\begin{equation}\label{eq:LR}
\begin{split}
Y = f(\bold{X}) = \beta_0 + \beta_1 X_1 + ... + \beta_i X_i + ... + \beta_N X_N + \epsilon,		
\end{split}
\end{equation}
where the intercept $\beta_0$ and the slopes $\beta_i$ are the unknown regression coefficients, and $\epsilon$ represents random observational error, usually assumed to be distributed normally with mean zero and some dispersion.

In our analysis, we use $\phi_{\rm sat}$ as the response variable $Y$, regressed against the 17 parameters in our parameter pool, as listed in Table~\ref{tb:predictors}. In short, we have 7 satellite-related, 6 central galaxy-related and 4 cluster-related quantities. 
When fitting the regression Eq.~\eqref{eq:LR}, we standardize the parameters $P_i$:
\begin{equation}\label{eq:normalization}
X_i = \frac{P_i - \mean{P}_i}{\sigma_{P_i}} \,,
\end{equation}
where $\mean{P}_i$ and $\sigma_{P_i}$ are the sample mean and standard deviation respectively of the parameter $P_i$, with the latter representing the width of the intrinsic scatter combined with measurement error.
Table~\ref{tb:parameter_mean_std} lists the $\mean{P}$ and $\sigma_{P}$ for our 17 parameters for DR8 redMaPPer satellites. 
The standardization means that even though our predictors $P_i$ have different magnitudes and
spreads, the fitted values $\beta_i$ all have essentially the same meaning: larger |$\beta_i$|
indicates a stronger effect of the regressor $X_i$ on $\phi_{\rm sat}$. We also quantify the level
of significance for any identified correlations using the $t$-value. The $t$-value, defined as the
ratio of $\beta_i$ to its standard error, can be positive or negative depending on the sign of
$\beta_i$. A larger |$t$| indicates a higher likelihood that $\beta_i \neq 0$, which means a
stronger relationship between $\phi_{\rm sat}$ and $X_i$. Statistically, under the assumption that
$\beta_i$ is asymptotically normal,\footnote{Suppose we do many measurements on $\beta_i$, with
  different but statistically similar sets of data, and plot the distribution of the resulting
  $\beta_i$. We say $\beta_i$ is asymptotically normal if the distribution is a gaussian at the mean
  of the true $\beta_i$ value with certain variance.} there is a direct link between the $t$-value
and $p$-value on the hypothesis test of whether $\beta_i=0$ or not. A $p$-value of 0.05 corresponds
to a 95\% confidence interval for $\beta_i$ that does not overlap with zero. 
We will count predictors with $p$-value < 0.05 as having significant effect on SA signal when doing model selection later. 

\subsubsection{Model Averaging}
\label{subsec:model_avg}

The next issue we need to face is how to choose a model with predictors that truly affect $\phi_{\rm
  sat}$. A linear model with all possible predictors has many free parameters to tune to fit the
data well, but may cause high variance. By contrast, a model with only few predictors is more
stable, but may underfit the data yielding a high bias. A good model results from achieving a balance between goodness of fit and complexity. With 17 predictor candidates in our parameter pool, there are a total of $2^{17}$ (= 131\,072) possible models.

In Paper I, we applied standard ``Forward Stepwise'' and ``Best Subset'' model selection methods to
identify a single best model, and interpreted our results based on that model alone. 
However, SA is a relatively weak signal, so identifying predictors that reflect the true underlying
physics rather than noise requires a careful treatment of model uncertainty, which
can lead to the selection of a model that by random chance includes uninformative predictors.
Thus in this work, we apply the ``model averaging'' technique (see e.g.~\citealt{Burnham10}
  and \citealt{Grueber11}, and references therein), which combines a portion of `good models' from
  the parent model pool by taking a weighted-average based on the performance scores of each model
  measured by some information criterion (such as AIC; \citealt{Akaike1998}). Through the model
  averaging process, fluctuations due to noise can be averaged out and the combined model is thus more stable.

 
The detailed model averaging analysis proceeds as follows:  
\begin{enumerate}
\item  As was the case in Paper I, we use best-subset selection to fit each of the $2^{17}$ possible models using {\tt R}'s 
        {\tt leaps} package, and for each we estimate the AIC:  
	\begin{equation}\label{eq:AIC}
	AIC_i = \frac{1}{n\hat{\sigma}_i^2} (RSS_i + 2 d_i \hat{\sigma}_i^2),
	\end{equation}
	where $RSS_i$ is the residual sum of squares, $d_i$ is the number of predictors used in the $i^{\rm th}$
	model, $\hat{\sigma}_i^2$ is an estimate of the variance of observational error $\epsilon$ shown in
	Eq.~\eqref{eq:LR}, and $n$ is the total number of satellite-central pairs. 
	A lower $AIC$ value indicates a better-fitting model.
\item  We convert each AIC value to a relative quantity: 
	\begin{equation}\label{eq:delta_AIC}
	\Delta AIC_i = AIC_i - {\rm min}(AIC),
	\end{equation}
	where ${\rm min}(AIC)$ represents the smallest $AIC$ value of all possible models in the previous step.
\item  Given the set of relative $AIC$ values, we compute the model-averaging weight for each model:
	\begin{equation}\label{eq:weight}
	w_i = \frac{\exp{(-\Delta AIC_i /2)}}{ \sum\limits_{k=1}^{2^{17}} \exp{(-\Delta AIC_k /2)}} \,.
	\end{equation}
\item  For computational efficiency, we apply a cut of $\Delta AIC=12$, meaning that we only average that subset of 
         485 models with $\Delta AIC_i < 12$. For this cut, the sum of weights is 0.996.
\item The final averaged regression estimate $\tilde{\beta}_j$ and variance estimate $\tilde{\sigma}^2_j$ for each of the 17 
         predictors are given by	
	\begin{eqnarray}
	\tilde{\beta}_j &=& \sum\limits_{i} w_i \beta_{j,i}  \bold{I}(\beta_{j} \in \mathcal{M}_i) \label{eq:betaj} \\
	\tilde{\sigma}^2_j &=& \sum\limits_{i} w_i [\sigma^2_{j,i} + (\beta_{j,i}-\tilde{\beta}_j)^2] \bold{I}(\beta_{j} \in \mathcal{M}_i) \,. \label{eq:var_betaj}
	\end{eqnarray}
	The first summation can be read as ``taking weighted average of the of the $j^{\rm th}$ regression coefficient over all 
	models with $\Delta AIC_i < 12$ where the $j^{\rm th}$ regressor appears'' (as indicated by the indicator function $\bold{I}$). 
	The second summation is the variance of the $j^{\rm th}$ regressor. If the estimated $\tilde{\beta}_j$ is more than 1.96 
	$\tilde{\sigma}_j$ away from zero, assuming $\tilde{\beta}_j$ is distributed normally with mean zero, this corresponds to a 
	$p$ value $<0.05$ that $\tilde{\beta}_j \neq 0$. One can interpret that the predictor $j$ is important for the SA signal.
\end{enumerate}
We note that the way in which $\tilde{\beta}_j$ is computed has the effect of shrinkage: the smaller number of models in 
which $\tilde{\beta}_j$ appears, the closer $\tilde{\beta}_j$ gets to zero. This method for computing $\tilde{\beta}_j$ is dubbed 
the zero method (\citealt{Burnham10}), which is contrasted against the natural average method, where we normalize the predictor 
estimate by dividing by $\sum\limits_{i} w_i \bold{I}(\beta_{j} \in \mathcal{M}_i)$. The zero method is preferable in situations
such as ours where the aim is to determine which predictors have the strongest effects on the response variable.

\subsection{Featured Predictor Selection -- re-Gaussianization shape}
\label{subsec:LR-reG}

In this subsection, we report the predictor selection result for our DR8 footprint sample with
$\phi_{\rm sat}$ measured using the re-Gaussianization method. All analyses below are properly weighted by $p_{\rm mem}$ for each satellite-central pair.

Table~\ref{tb:DR8_model_list} lists those models that have $\Delta AIC < 3$, with the first row containing the model with the
smallest $AIC$ value, i.e., the model that would be selected in a traditional implementation of best-subset selection using $AIC$
as the fit metric (or equivalently using Mallow's $C_p$, which is proportional to $AIC$).
The first column indicates the total number of regressors included in a model, and 
the subsequent 2nd to the 18th columns records the regression coefficients $\beta_j$ formulated in Eq.~\eqref{eq:LR}. 
%
%
Some predictors, such as log($r$/$R_{\rm 200m}$) and satellite luminosity $^{0.1}Mr$, appear in all of the top models, while
less important predictors appear only occasionally. 
In Table~\ref{tb:DR8reG_Mavg} we lists the values of $\tilde{\beta}$ and $\tilde{\sigma}$, as well as the absolute $t$ value, for 
each predictor after averaging over all models with $\Delta AIC < 12$ (485 models in total; see eqs.~\eqref{eq:betaj} and ~\eqref{eq:var_betaj}). Predictors with $|t|>1.96$ are identified as significant in affecting SA.
Under the re-Gaussianization shape measurement, we identify log(r/R$\rm_{200m}$), $^{0.1}$Mr, satellite ellipticity e$_{\rm sat}$, 
and $fracDeV$ as significant predictors. To reiterate a point made above,
in this work we are not interested in using the $\tilde{\beta}$ values to predict $\phi_{\rm sat}$;
our interest lies in quantifying the significances of the predictors (as indicated by $t$ values) and in exhibiting their effect on $\phi_{\rm sat}$ (as indicated in the sign of $\tilde{\beta}$).

\begin{figure*}
\begin{center}
\includegraphics[trim=0.15cm 0cm 1.0cm 0cm,width=0.8\textwidth]{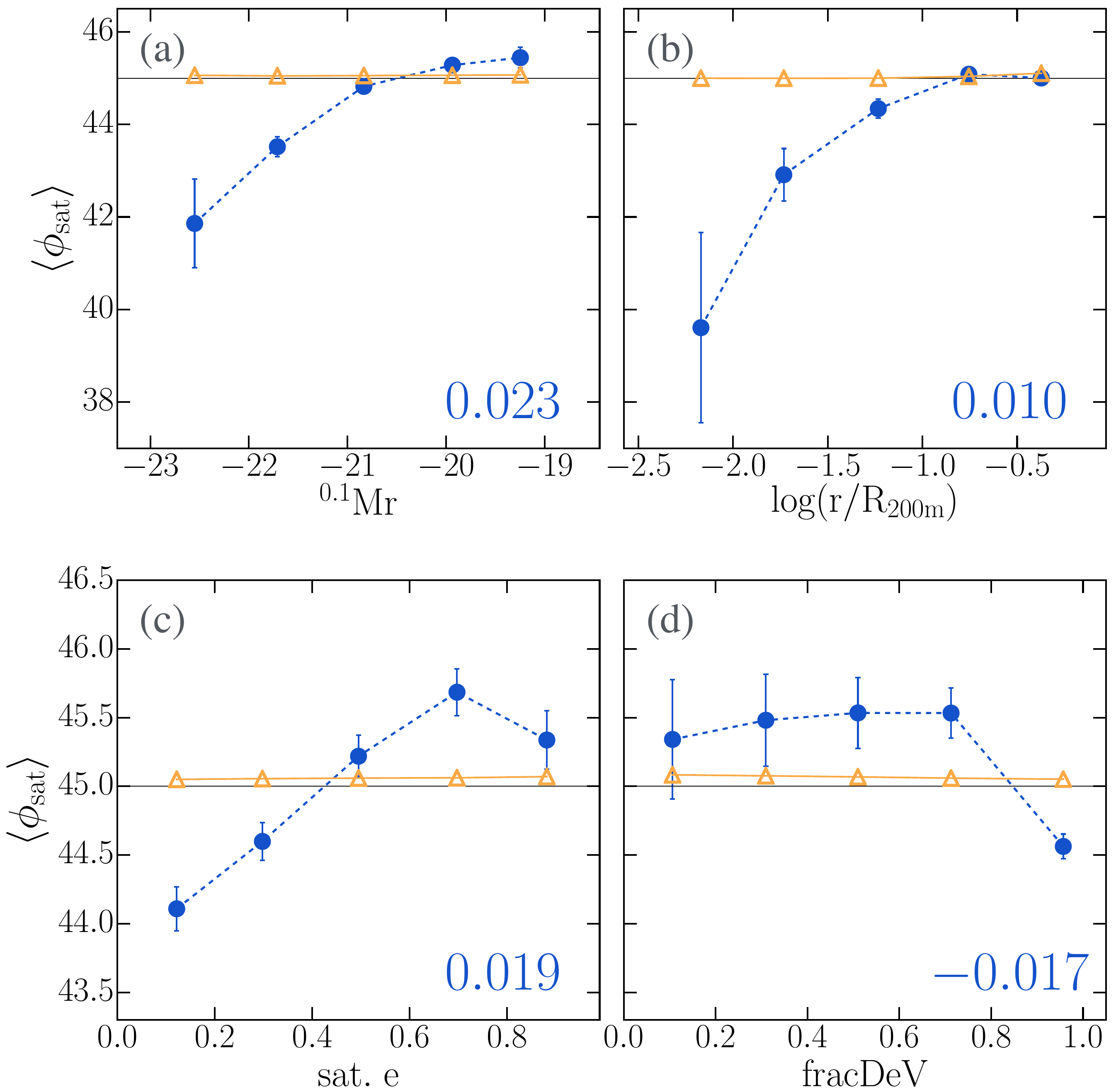}
\caption{Averaged satellite alignment angle $\mean{\phi_{\rm sat}}$ of redMaPPer member galaxies in
  the SDSS DR8 footprint sample as a function of the 4 significant predictors whose $|t|>1.96$ shown in
  Table~\ref{tb:DR8reG_Mavg}. The correlation coefficient between $\phi_{\rm sat}$ and each
  predictor is labeled on the lower-right corner of each panel. 
\textit{One can see that $\mean{\phi_{\rm sat}}$ < 45$^{\circ}$ for satellites that have higher
  luminosity, are located closer to cluster center, are rounder in shape, and have a higher $fracDeV$. 
  Especially for the subsamples of luminous satellites, we detect a very significant SA signal.}
The triangular orange markers show the estimated level of lensing contamination from background
galaxies that are wrongly included in the SA analysis (see Appendix~\ref{app:est_len} for
details), resulting in $\mean{\phi_{\rm sat}}\sim$ 45.1$^\circ$ across all 
bins. 
}
\label{fig:PhiSAT_x_DR8}
\end{center}
\end{figure*}

Fig.~\ref{fig:PhiSAT_x_DR8} illustrates these trends by plotting the
averaged value of $\phi_{\rm sat}$ in bins of each selected featured predictor, with the correlation
coefficient between $\phi_{\rm sat}$ and each predictor also provided. 
Clearly the satellite luminosity ($^{0.1}Mr$) and separation from the central galaxy in units of
$R_{\rm 200m}$ (log($r$/R$_{\rm 200m}$)) are prominent predictors for satellite alignments.  
There are sub-populations with measured $\mean{\phi_{\rm sat}}>45^\circ$, although with less than
3$\sigma$ significance, for example at faint luminosity or low $fracDeV$. This may in part be due to
lensing contamination from background galaxies that are wrongly included in the satellite
sample. The orange triangular points shown in Fig.~\ref{fig:PhiSAT_x_DR8} are the estimates of
lensing contamination based on the assumption that the $p_{\rm mem}$ values accurately reflect
reality; we refer readers to Appendix~\ref{app:est_len} for details. The estimated contribution from
lensing contamination to $\mean{\phi_{\rm sat}}\approx 45.1^\circ$ across all subsamples. 
 
In Sec.~\ref{subsec:DR8_phi_sat} and~\ref{subsec:e+}, no SA signal was detected based on
re-Gaussianization shape measurement, when averaging over all satellite-central pairs.  The results
in this section demonstrate that there do exist statistically significant SA effects for certain
subsamples of satellites, such as those that are intrinsically brighter or located closer to their
host central galaxies. 
We further discuss these selected predictors in Sec.~\ref{sec:discussion}. 

\begin{landscape}

\begin{table}
\caption{Mean and standard deviation values (weighted by $p_{\rm mem}$) of the 17 parameters for our DR8 $p_{\rm mem}>0.55$ redMaPPer satellites. 
}

\begin{tabular}{lccccccccccccccccc}
\hline 

  & log(r/R$\rm_{200m}$) & $^{0.1}$Mr & color & e$_{\rm sat}$ & R$_{\rm eff, sat}$  & $\theta_{\rm cen}$ & fDeV & dom & $^{0.1}$Mr$_{\rm cen}$ & color$_{\rm cen}$ & e$_{\rm cen}$ & R$_{\rm eff, cen}$ & P$_{\rm cen}$ & $\lambda$ & z & e$\rm_{cluster}$ & $\Delta_{\rm R}$ \\ \hline  \hline

mean & -0.68 & -20.40 & 0.89 & 0.45 & 1.32  &	42.09 & 0.77 & -0.89 & -22.35 & 0.98 & 0.25 &  0.02 & 0.87 & 41.99	& 0.24 & 0.20 & 0.01 \\
std. 	 &   0.34 &  0.76   & 0.12 & 0.26 & 0.20  & 26.09 & 0.30 & 0.60  &    0.49 & 0.07 & 0.16 & 	0.15 & 0.17 & 25.01  & 0.07 & 0.11 & 0.11 \\

\hline

\label{tb:parameter_mean_std}
\end{tabular}
\end{table}

\begin{table}
\caption{
A subset of top regression models ranked by the goodness of performance for DR8 re-Gaussianization shape measurements.
The 1st column is the total number of predictors involved in a model, columns $2\sim18$ show the regression coefficients for our 17 predictors listed in the same order 
as that of Table~\ref{tb:predictors}. The 2nd last column lists the $\Delta AIC$ values, and the last column shows the weighting factor for each model that we applied when doing model averaging. 
}
\begin{tabular}{cccccccccccccccccccc}
\hline 

Np  & log(r/R$\rm_{200m}$) & $^{0.1}$Mr & color & e$_{\rm sat}$ & R$_{\rm eff, sat}$  & $\theta_{\rm cen}$ & fDeV & dom & $^{0.1}$Mr$_{\rm cen}$ & color$_{\rm cen}$ & e$_{\rm cen}$ & R$_{\rm eff, cen}$ & P$_{\rm cen}$ & $\lambda$ & z & e$\rm_{cluster}$ & $\Delta_{\rm R}$ & $\Delta$AIC & weight \\ \hline  \hline

6  & 0.32  & 0.49  &   & 0.28  & -0.14  &   & -0.28  &   & -0.16  &   &   &   &   &   &   &   &   & 0  &  0.0296  \\
7  & 0.32  & 0.49  &   & 0.28  & -0.14  & 0.08  & -0.28  &   & -0.16  &   &   &   &   &   &   &   &   & 0.82  &  0.0196  \\
7  & 0.32  & 0.48  &   & 0.29  & -0.14  &   & -0.28  &   & -0.16  &   &   & 0.06  &   &   &   &   &   & 1.49  &  0.014  \\
7  & 0.31  & 0.47  &   & 0.29  & -0.14  &   & -0.3  &   & -0.17  &   &   &   &   &   & -0.06  &   &   & 1.52  &  0.0138  \\
7  & 0.32  & 0.49  &   & 0.28  & -0.14  &   & -0.28  &   & -0.16  &   & 0.05  &   &   &   &   &   &   & 1.64  &  0.013  \\
7  & 0.32  & 0.49  &   & 0.28  & -0.14  &   & -0.28  & -0.05  & -0.13  &   &   &   &   &   &   &   &   & 1.81  &  0.012  \\
7  & 0.32  & 0.49  &   & 0.28  & -0.14  &   & -0.28  &   & -0.17  & -0.04  &   &   &   &   &   &   &   & 1.83  &  0.0118  \\
5  & 0.31  & 0.5  &   & 0.32  &   &   & -0.23  &   & -0.17  &   &   &   &   &   &   &   &   & 1.9  &  0.0115  \\
6  & 0.32  & 0.48  &   & 0.28  & -0.14  &   & -0.28  & -0.14  &   &   &   &   &   &   &   &   &   & 2.01  &  0.0108  \\
7  & 0.32  & 0.49  &   & 0.28  & -0.14  &   & -0.28  &   & -0.16  &   &   &   &   &   &   & -0.02  &   & 2.15  &  0.0101  \\
7  & 0.32  & 0.49  & -0.02  & 0.28  & -0.14  &   & -0.27  &   & -0.16  &   &   &   &   &   &   &   &   & 2.22  &  0.0098  \\
7  & 0.32  & 0.49  &   & 0.28  & -0.14  &   & -0.28  &   & -0.16  &   &   &   &   &   &   &   & 0.01  & 2.24  &  0.0097  \\
8  & 0.32  & 0.48  &   & 0.28  & -0.14  & 0.08  & -0.28  &   & -0.16  &   &   & 0.06  &   &   &   &   &   & 2.25  &  0.0096  \\
7  & 0.32  & 0.49  &   & 0.28  & -0.14  &   & -0.28  &   & -0.16  &   &   &   & 0.01  &   &   &   &   & 2.26  &  0.0096  \\
7  & 0.32  & 0.49  &   & 0.28  & -0.14  &   & -0.28  &   & -0.16  &   &   &   &   & 0  &   &   &   & 2.27  &  0.0095  \\
8  & 0.31  & 0.47  &   & 0.29  & -0.14  & 0.08  & -0.29  &   & -0.17  &   &   &   &   &   & -0.06  &   &   & 2.31  &  0.0093  \\
8  & 0.32  & 0.49  &   & 0.28  & -0.14  & 0.08  & -0.28  &   & -0.16  &   & 0.05  &   &   &   &   &   &   & 2.44  &  0.0087  \\
8  & 0.32  & 0.49  &   & 0.28  & -0.14  & 0.08  & -0.28  & -0.06  & -0.13  &   &   &   &   &   &   &   &   & 2.61  &  0.008  \\
8  & 0.32  & 0.49  &   & 0.28  & -0.14  & 0.07  & -0.27  &   & -0.17  & -0.04  &   &   &   &   &   &   &   & 2.67  &  0.0078  \\
7  & 0.32  & 0.48  &   & 0.28  & -0.14  &   & -0.28  & -0.15  &   &   &   &   &   & 0.08  &   &   &   & 2.68  &  0.0078  \\
6  & 0.31  & 0.5  &   & 0.31  &   & 0.07  & -0.23  &   & -0.17  &   &   &   &   &   &   &   &   & 2.76  &  0.0075  \\
7  & 0.32  & 0.48  &   & 0.28  & -0.14  & 0.08  & -0.28  & -0.14  &   &   &   &   &   &   &   &   &   & 2.83  &  0.0072  \\

\hline

\label{tb:DR8_model_list}
\end{tabular}
\end{table}

\end{landscape}

\begin{landscape}

\begin{table}
\caption{
Model averaged result for our DR8 footprint sample (see Table~\ref{tb:samples}) based on re-Gaussianization shape measurements.
The 1st row shows the estimated regression coefficient based on Eq.\eqref{eq:betaj}. The 2nd row lists the standard error of $\tilde{\beta}$. 
The 3rd row is the absolute value of the $t$ value $=|\tilde{\beta}/\tilde{\sigma}|$. Parameters with $|t|>1.96$ is identified as important predictors in affecting SA effect.
Here we find that log(r/R$\rm_{200m}$),  $^{0.1}$Mr, satellite ellipticity, and $fracDeV$ are selected as significant predictors. 
}
\begin{tabular}{lccccccccccccccccc}
\hline 
  & log(r/R$\rm_{200m}$) & $^{0.1}$Mr & color & e$_{\rm sat}$ & R$_{\rm eff,sat}$  & $\theta_{\rm cen}$ & fDeV & dom & $^{0.1}$Mr$_{\rm cen}$ & color$_{\rm cen}$ & e$_{\rm cen}$ & R$_{\rm eff, cen}$ & P$_{\rm cen}$ & $\lambda$ & z & e$\rm_{cluster}$ & $\Delta_{\rm R}$ \\ \hline  \hline

$\tilde{\beta}$ &
0.314  & 0.48  & -0.003  & 0.289  & -0.115  & 0.032  & -0.273  & -0.028  & -0.133  & -0.008  & 0.011  & 0.012  & 0.001  & 0.005  & -0.016  & -0.003  & 0.001  \\
$\tilde{\sigma}$ &
0.063  & 0.07  & 0.026  & 0.068  & 0.065  & 0.049  & 0.074  & 0.059  & 0.069  & 0.032  & 0.035  & 0.035  & 0.023  & 0.03  & 0.043  & 0.024  & 0.02 \\ 
$|t|$ &
4.982  & 6.892  & 0.111  & 4.257  & 1.781  & 0.639  & 3.71  & 0.469  & 1.932  & 0.258  & 0.317  & 0.325  & 0.037  & 0.152  & 0.377  & 0.123  & 0.052 \\
\hline
\label{tb:DR8reG_Mavg}
\end{tabular}
\end{table}

\begin{table}
\caption{
Model averaged result for our DR7 footprint sample (see Table~\ref{tb:samples}) based on de Vaucouleurs shape measurements.
The selected significant predictors ($|t|>1.96$) are: log(r/R$\rm_{200m}$),  $^{0.1}$Mr, satellite ellipticity, $\theta_{\rm cen}$, and $fracDeV$. 
}
\begin{tabular}{lccccccccccccccccc}
\hline 
  & log(r/R$\rm_{200m}$) & $^{0.1}$Mr & color & e$_{\rm sat}$ & R$_{\rm eff,sat}$  & $\theta_{\rm cen}$ & fDeV & dom & $^{0.1}$Mr$_{\rm cen}$ & color$_{\rm cen}$ & e$_{\rm cen}$ & R$_{\rm eff, cen}$ & P$_{\rm cen}$ & $\lambda$ & z & e$\rm_{cluster}$ & $\Delta_{\rm R}$ \\ \hline  \hline

$\tilde{\beta}$ &
0.628  & 0.617  & -0.021  & 0.41  & -0.02  & 0.229  & -0.218  & -0.01  & -0.094  & -0.035  & 0.016  & 0.024  & 0.046  & 0.065  & 0.001  & 0  & -0.003 \\
$\tilde{\sigma}$ &
0.066  & 0.072  & 0.046  & 0.07  & 0.046  & 0.065  & 0.073  & 0.044  & 0.074  & 0.054  & 0.041  & 0.048  & 0.062  & 0.067  & 0.021  & 0.016  & 0.022 \\
$|t|$ &
9.536  & 8.62  & 0.448  & 5.886  & 0.433  & 3.506  & 2.977  & 0.227  & 1.268  & 0.657  & 0.397  & 0.497  & 0.742  & 0.979  & 0.069  & 0.002  & 0.133 \\
\hline
\label{tb:DR7deV_Mavg}
\end{tabular}
\end{table}

\begin{table}
\caption{
Similar to Table~\ref{tb:DR7deV_Mavg} but for results based on isophotal shape measurements. 
The selected significant predictors ($|t|>1.96$) are: log(r/R$\rm_{200m}$),  $^{0.1}$Mr, satellite ellipticity, $\theta_{\rm cen}$, $fracDeV$, and redshift. 
}
\begin{tabular}{lccccccccccccccccc}
\hline 
  & log(r/R$\rm_{200m}$) & $^{0.1}$Mr & color & e$_{\rm sat}$ & R$_{\rm eff,sat}$  & $\theta_{\rm cen}$ & fDeV & dom & $^{0.1}$Mr$_{\rm cen}$ & color$_{\rm cen}$ & e$_{\rm cen}$ & R$_{\rm eff, cen}$ & P$_{\rm cen}$ & $\lambda$ & z & e$\rm_{cluster}$ & $\Delta_{\rm R}$ \\ \hline  \hline

$\tilde{\beta}$ &
0.953  & 0.713  & -0.154  & 0.451  & 0.003  & 0.371  & -0.325  & 0.014  & -0.022  & 0.002  & 0.003  & 0.001  & 0.107  & 0.055  & 0.177  & -0.052  & -0.008 \\
$\tilde{\sigma}$ &
0.067  & 0.079  & 0.071  & 0.069  & 0.03  & 0.066  & 0.076  & 0.051  & 0.056  & 0.028  & 0.028  & 0.028  & 0.068  & 0.062  & 0.076  & 0.06  & 0.033 \\
$|t|$ &
14.282  & 9.074  & 2.163  & 6.498  & 0.093  & 5.629  & 4.301  & 0.279  & 0.397  & 0.063  & 0.105  & 0.03  & 1.57  & 0.895  & 2.32  & 0.868  & 0.239 \\
\hline
\label{tb:DR7iso_Mavg}
\end{tabular}
\end{table}

\end{landscape}

\subsection{Featured Predictor Selection -- \ \ \ \ \ \ \ \ \ \ \ \ \ \ \ \ \ \ de Vaucouleurs and isophotal shapes}

Here we repeat the predictor selection process as in the previous section, now using the DR7 sample
satellites that have $\phi_{\rm sat}$ well-measured using all three shape measurement methods. Using
both de Vaucouleurs and isophotal shapes result in a nonzero net SA signal detection in the overall
sample as shown in Sec.~\ref{subsec:DR7_phi_sat}. It is therefore interesting to check if the
predictor selection result is consistent with that based on re-Gaussianization shape, where the
detected SA signal is small. If we select a different set of predictors, we must consider whether
they are caused by a fake systematic alignment signal captured in de Vaucouleurs and isophotal
shapes, or they could be physically reasonable predictors that are authentically associated with the SA phenomenon, but are not selected
out in the re-Gaussianization shape due to its sensitivity to different regions of the galaxy light profiles.
 
We summarize our predictor selection results for de Vaucouleurs and isophotal shapes in Tables~\ref{tb:DR7deV_Mavg} and ~\ref{tb:DR7iso_Mavg}. 
Under the criterion of $|t|>1.96$, for de Vaucouleurs shape, in addition to the predictors that have been identified based on re-Gaussianization shape (Table~\ref{tb:DR8reG_Mavg}), one extra predictor, $\theta_{\rm cen}$, is added with a fairly high $t$-value. 
For isophotal shape, we identify two new predictors compared to those from re-Gaussianization shapes: $\theta_{\rm cen}$ and redshift. 
As revealed in the sign of the $\tilde{\beta}$, satellites with smaller $\theta_{\rm cen}$ (stronger
central galaxy alignment with the shape of the satellite galaxy distribution) and smaller redshift have stronger isophotal SA signal.

\begin{figure*}
\begin{center}
\includegraphics[trim=0.15cm 0cm 1.0cm 0cm,width=0.95\textwidth]{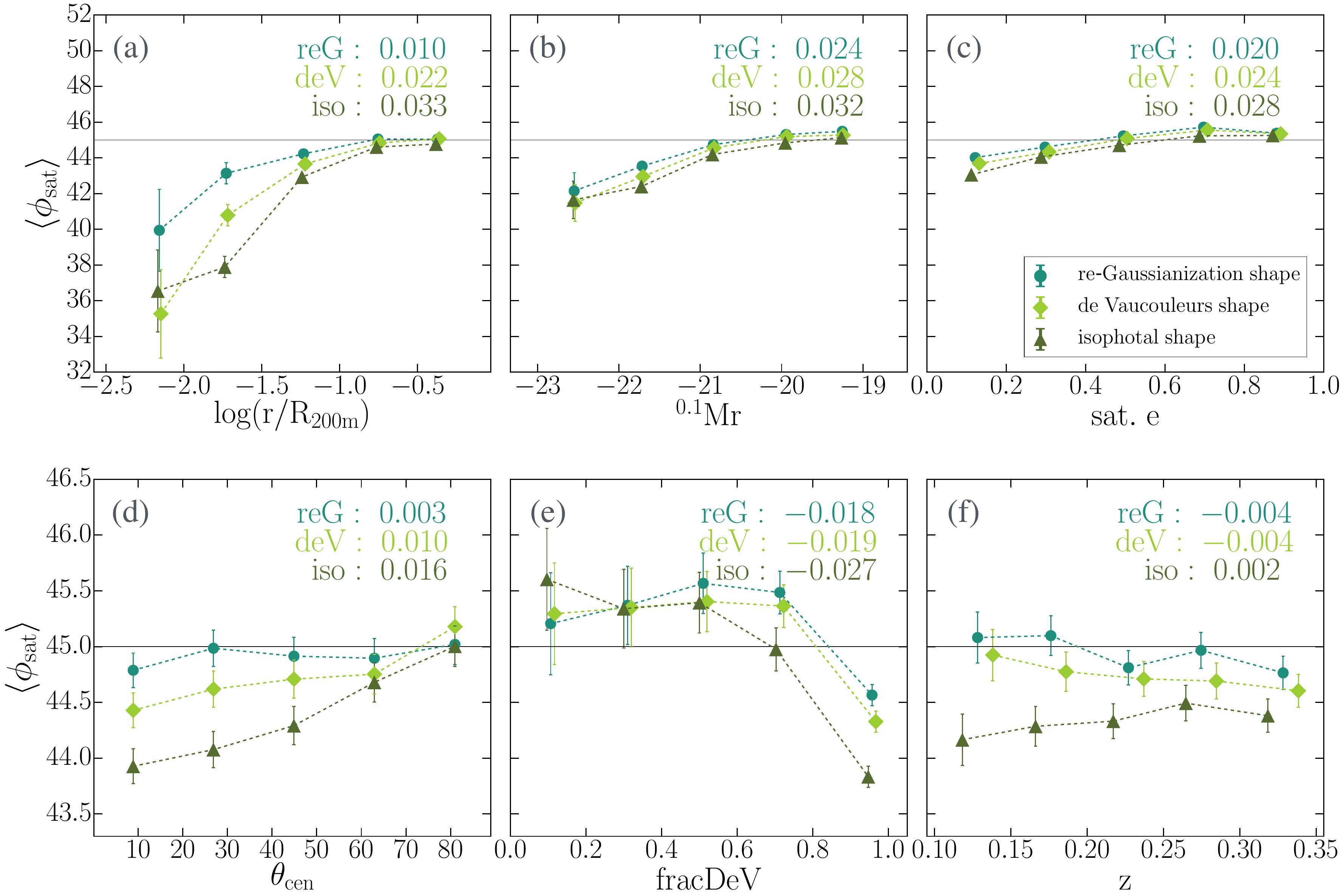}
\caption{Averaged satellite alignment angle $\mean{\phi_{\rm sat}}$ of redMaPPer members in the SDSS DR7
  footprint as a function of the 6 significant predictors whose $|t|>1.96$ as shown in
  Tables~\ref{tb:DR7deV_Mavg} and ~\ref{tb:DR7iso_Mavg}. Correlation coefficients between $\phi_{\rm sat}$ and $x$-axes parameters are shown in the upper right corner of each
  panel.
}
\label{fig:PhiSAT_x_DR7}
\end{center}
\end{figure*}

Fig.~\ref{fig:PhiSAT_x_DR7} shows the averaged SA angle $\mean{\phi_{\rm sat}}$ of our DR7 footprint
sample in bins of the identified predictors. 
The correlation coefficients measured between $\phi_{\rm sat}$ using the three shape measurements and each predictor are shown in the legend. 
The correlations become tighter as we move from re-Gaussianization to isophotal shapes for most
predictors -- log($r$/R$_{\rm 200m}$), satellite $^{0.1}Mr$, satellite ellipticity, $\theta_{\rm
  cen}$, and  $fracDeV$. However, the correlation coefficient for the redshift has a different sign
when measured in isophotal shape vs.\ the other two shapes.  We will discuss the redshift dependence
in more detail in Sec.~\ref{sec:discussion}.

In general, for all differences between methods, we must consider whether they originate from
systematic effects, or from real differences between the methods due to their sensitivity to
different parts of the galaxy light profiles and isophotal twisting.

%% file: 5_cpShapes.tex
\section{Origin of discrepancy in detected Satellite Alignment signal with different shape measurement methods}
\label{sec:cp_shape}

As reported in Fig.~\ref{fig:phisat_dist_DR7}, the detected SA signal strength depends on shape measurement methods.
The isophotal shape detects the strongest SA signal, followed by de Vaucouleurs shape then
re-Gaussianization shape. This trend is consistent with the large-scale IA measurement done by
\citet{Singh16} using these three shape measurement algorithms.

In this section, we discuss possible reasons for this discrepancy. 
We note that in cluster systems, the difference in SA angle ($|\phi_{\rm sat, iso}-\phi_{\rm sat,
  reG}|$) is identical to the PA difference, $|\rm{PA_{iso}}-\rm{PA_{reG}}|$ measured in the RA-dec
frame.  We can therefore use  $|\rm{PA_{iso}}-\rm{PA_{reG}}|$ for our tests, which allows us to
introduce galaxies that do not belong to any cluster for tests of statistical and systematic errors.

\begin{figure} 
\begin{center}
\includegraphics[width=0.45\textwidth]{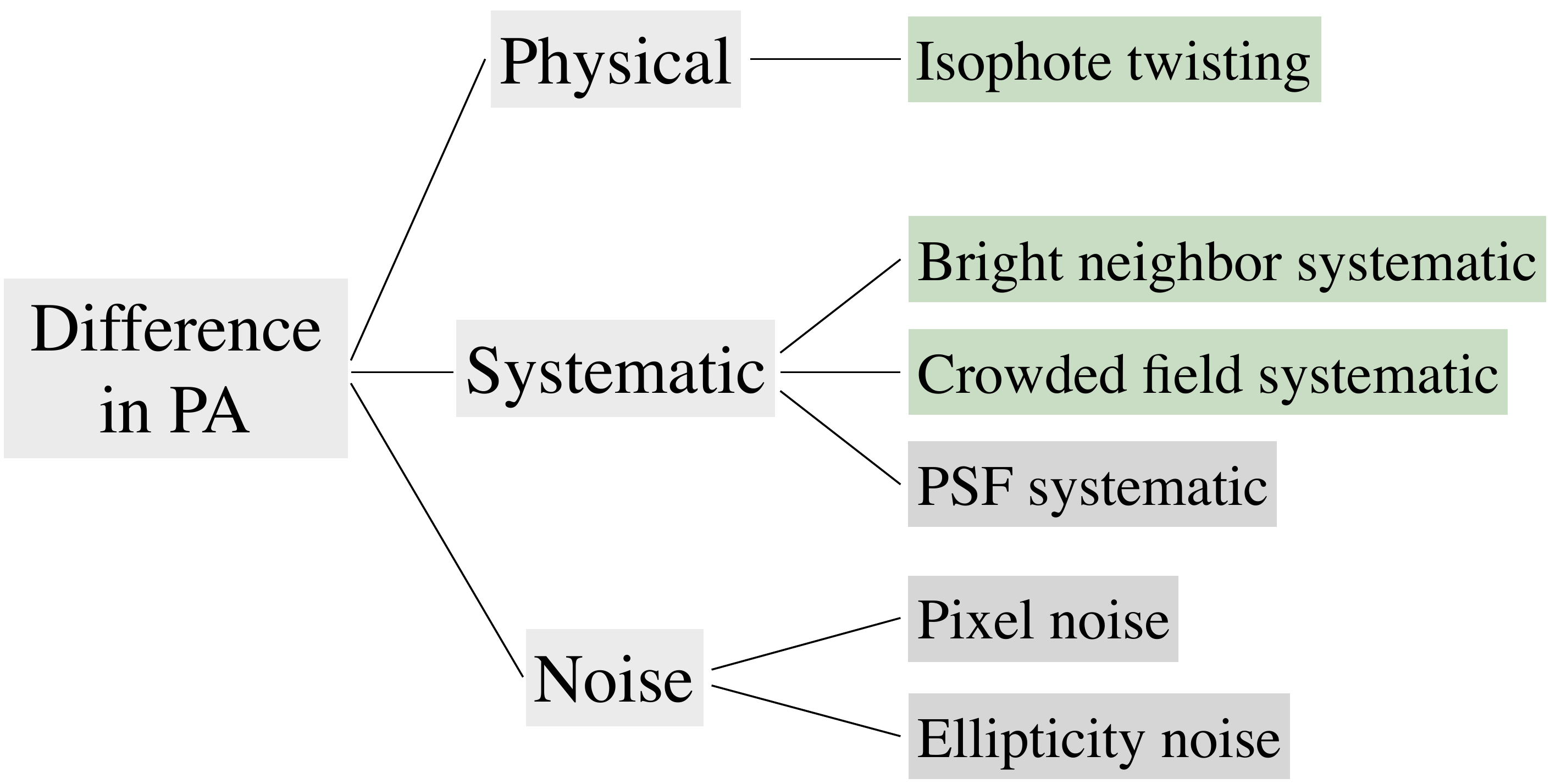}
\caption{The origin of discrepancies in PA differences measured with different shape measurement
  methods. 
  Three origins for these discrepancies are explored: discrepancies due real physical isophote
  twisting effect (Sec.~\ref{sec: physical}), discrepancies caused by systematic errors, and
  discrepancies due to noise. The systematic origins can be further classified into three factors:
  the bright neighbor and crowded field systematics (Sec.~\ref{sec: cluster field systematics}) and
  the PSF systematic (Sec.~\ref{sec: psf systematic}). The origins of noise are separated into two
  sources: pixel (Sec.~\ref{sec: pixel noise}) and ellipticity noises (Sec.~\ref{sec: e noise}). The
  items colored in green background are factors that especially important for galaxies residing in
  cluster fields, while the ones colored in grey background are relevant in all possible environments. 
  }
\label{fig:PA_difference}
\end{center}
\end{figure}

In what follows, we classify the origin of the discrepancy in the measured galaxy PA into three dominant factors, as summarized in Fig.~\ref{fig:PA_difference}. 
Our high-level goal is to consider all possible factors that may contribute to the measured
difference in average SA angle for our redMaPPer cluster sample, and determine whether the result is dominated by noises and systematics, or we do detect any interesting physical effects.

\subsection{PA discrepancies due to noise}
\label{sec: noise}

The first factor that affects the difference in PA is simply noise. 
We further separate the origin of noise into ellipticity noise and pixel noise.

\subsubsection{Ellipticity noise}
\label{sec: e noise}
The ellipticity noise is perhaps the most dominant factor in determining how precisely the PAs are measured. 
For a given $S/N$ ratio of a detection, the uncertainty in the PA is larger for rounder galaxy (see Table 1 of \citealt{Refregier12}). 
We demonstrate this effect in Fig.~\ref{fig:PAdiff_emrz}a using our non-cluster field sample, as defined in Sec.~\ref{subsec:sample2}. 
We plot the PA differences between isophotal and re-Gaussianization shapes as a function of the
ellipticity based on re-Gaussianization measurement. The filled circle shows the mean of the
$|\rm{PA_{iso}}-\rm{PA_{reG}}|$ value in ellipticity bins, revealing that the averaged differences in PA become larger when galaxies are rounder. 

\subsubsection{Pixel noise}
\label{sec: pixel noise}

Pixel noise arises from the Poisson noise in the sky and object flux in CCD measurements. 
Its impact is strongest on the images of faint galaxies, which have relatively low signal but still
experience all the Poisson noise due to the background level.
Pixel noise makes it difficult to measure the shape of low $S/N$ galaxies,
especially when those galaxies are also poorly resolved compared to the PSF \citep[e.g.,][]{Refregier12}.
As a consequence, there may also be an apparent redshift trend.

\begin{figure*}
\begin{center}
\includegraphics[trim=0.15cm 0cm 1.0cm 0cm,width=0.85\textwidth]{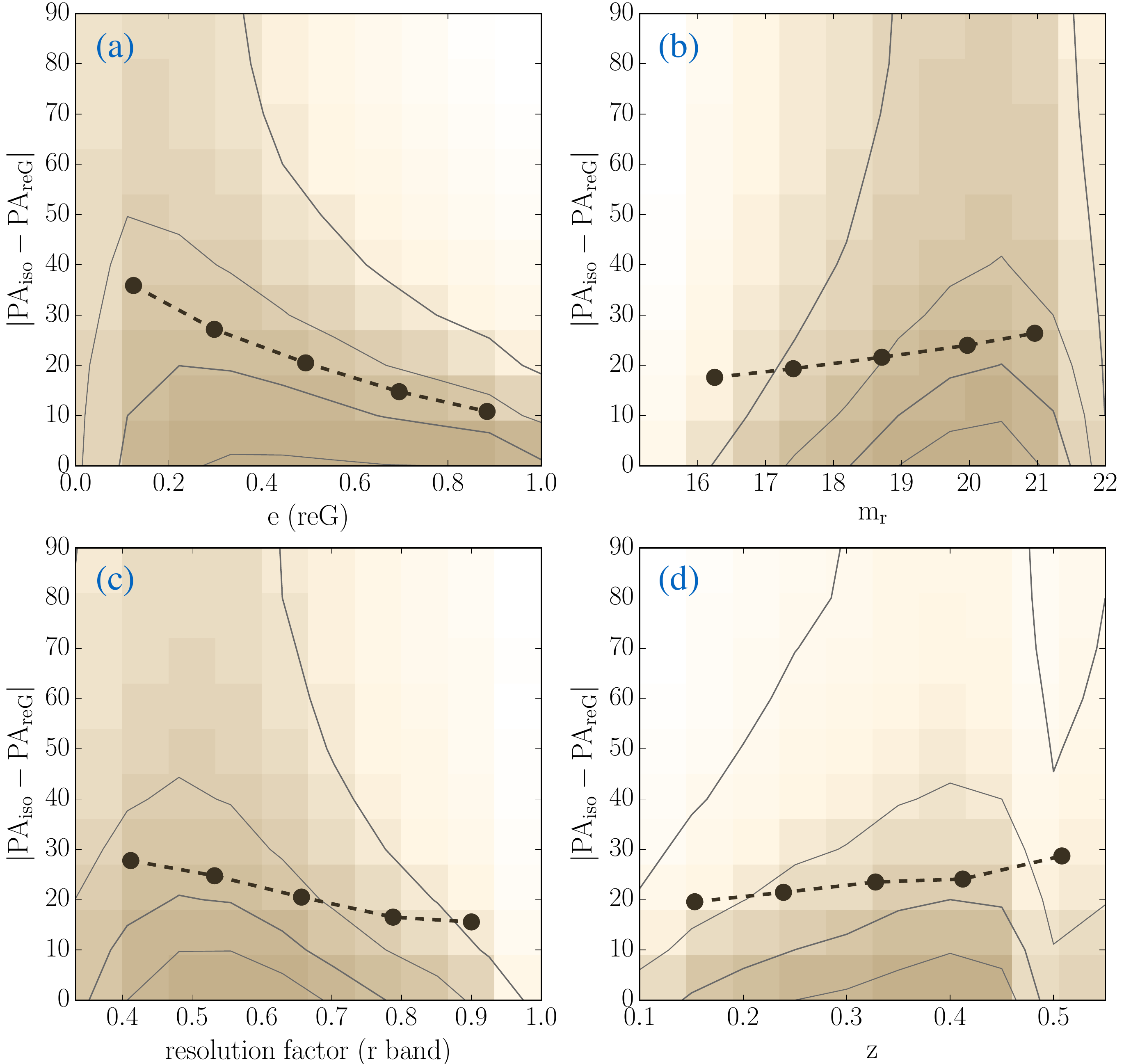}
\caption{Absolute PA differences between isophotal and re-Gaussianization shapes as a function of galaxy
  ellipticity, mag $r$, r-band resolution factor, and redshift. The grey contour levels indicate the
  level below which 20\%, 45\%, 70\%, and 95\% of the central-satellite pairs are located. The
  filled solid dots indicate the averaged $|\rm{PA_{iso}}-\rm{PA_{reG}}|$ values in bins of the
  horizontal axes values. \textit{As shown, the averaged differences in PA increase for galaxies
    that are rounder in shape, with fainter $m_r$, lower resolution, and located at higher $z$.}}
\label{fig:PAdiff_emrz}
\end{center}
\end{figure*}

Figs.~\ref{fig:PAdiff_emrz}b,~\ref{fig:PAdiff_emrz}c, and~\ref{fig:PAdiff_emrz}d show the scatter plots for 
$|\rm{PA_{iso}}-\rm{PA_{reG}}|$ as a function of apparent $r$-band magnitude, r-band resolution
factor\footnote{The resolution factor reflects how resolved a galaxy is compared to its PSF, with 0
  (1) indicating a completely unresolved (perfectly well-resolved) galaxy. See
  Appendix A of \citet{Reyes12} for its definition.}, and photo-$z$ respectively from the non-cluster field sample. 
We can see that the averaged differences in PA (filled circle) go up for galaxies with fainter
$m_r$, lower resolution, and higher $z$ as expected.

\subsection{PA discrepancies due to systematic errors}
\label{sec: systematic}
Systematic errors due to data analysis methods or contamination from nearby objects (see
Fig.~\ref{fig:PA_difference}) may also cause PA discrepancies between different methods. 

\subsubsection{PSF systematics}
\label{sec: psf systematic}
Different shape measurements correct for the effects of the PSF on galaxy images at different levels. 
If not removed properly, the residuals of PSF would bias the resulting shape parameters: 
In the case that PSFs are round and galaxies have elliptical isophotes, on average, the PA will be
unaffected by the PSF convolution, with only the magnitude of the galaxy shape being affected. If PSFs are elongated toward specific directions, then both the measured ellipticity and PA of galaxies are contaminated.

Isophotal shapes do not explicitly correct for the PSF convolution; de Vaucouleurs shapes partially
correct for the PSF using an approximate (double-Gaussian) PSF model. The re-Gaussianization shapes consider a full PSF model to remove PSF systematics in order to recover small weak lensing signals. 
Therefore, part of the discrepancy in PA measured based on different shape measurements may be due to
different levels of PSF anisotropy removal.

\citet{Singh16} explored the additive bias in galaxy shape measurements due to residual PSF
anisotropy, resulting in a coherent additive bias in the measured shapes of SDSS LOWZ galaxies using
these three shape measurements.
They found that for both re-Gaussianization and isophotal shapes, their additive biases are quite small. 
However, for de Vaucouleurs shapes, the additive bias is about a factor of 10 larger than with the other two shape measurements (see their Fig.~5).
They claimed that this may be due to the fact that the de Vaucouleurs modeling uses an approximate
PSF model.  If those results are relevant for galaxies in cluster fields, then part of the PA discrepancy 
between de Vaucouleurs shape and the other two is contributed by the
residual PSF anisotropy in the galaxy shapes.  In fact, they should be even more important here,
because the galaxies used in our analysis are smaller and fainter, resulting in a greater
susceptibility to PSF anisotropy modeling errors.

\subsubsection{Bright neighbor and crowded field systematics}
\label{sec: cluster field systematics}

Different shape measurement methods determine galaxy PA based on different parts of the galaxy's light profile. The re-Gaussianization method puts more emphasis on the central region of a galaxy's profile. The de Vaucouleurs shape includes both central and outer extended wings of the light profile to fit PA, while the isophotal shape traces the outermost region of a galaxy along the 25 mag/arcsec$^2$ isophote. 
These choices may make the latter two methods more sensitive to artifacts in the de-blending and sky subtraction processes, leading to spurious SA signals.
The two dominant systematics that affect the de-blending and sky subtraction processes and further
contribute to the PA discrepancy are bright neighbor and crowded field
systematics. 
The bright neighbor systematic arises due to the contamination of light from nearby bright neighbors in the galaxy for which
we are attempting to measure a shape. 
In cluster-like high density regions, the measured galaxy PA could be biased coherently pointing
toward the high-density direction due to the intracluster light or due to the fact that the large
number of bright galaxies causes a misestimated sky gradient. We refer this second effect as the crowded field systematic.

\begin{figure*}
\begin{center}
\includegraphics[trim=0.15cm 0cm 1.0cm 0cm,width=0.9\textwidth]{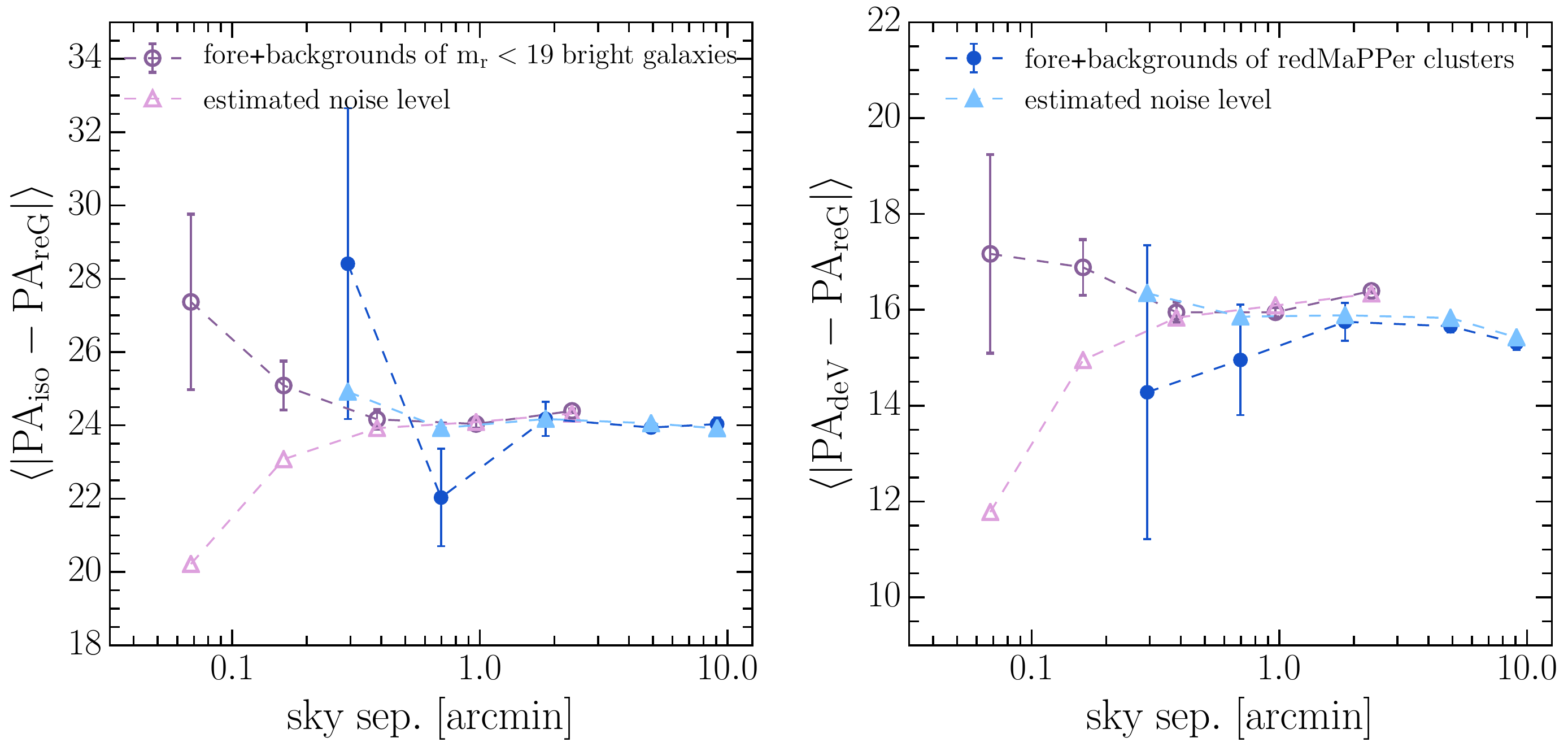}
\caption{Left panel: mean absolute PA differences (MAPAD) between isophotal and re-Gaussianization shapes as a function of
  projected angular separation on the sky for galaxies that are in the foreground and background of
  $m_r$ <19 non-cluster field bright galaxies (purple open circle), and for galaxies in the foreground and background of redMaPPer clusters (dark blue filled circle). The lighter color points shown with triangular symbols indicate the estimated $\mean{\rm{|PA_{iso}-PA_{reG}|}}$ contribution from noise (see the text for details). 
Right panel: similar to the left panel but for the MAPAD between de Vaucouleurs and re-Gaussianization shapes, $\mean{\rm{|PA_{deV}-PA_{reG}|}}$.
\textit{As shown, the MAPAD increases toward small sky separation.}
  }
\label{fig:PAdiff_skysep}
\end{center}
\end{figure*}

Below we start by estimating the level of bright neighbor systematic.
In the left panel of Fig.~\ref{fig:PAdiff_skysep}, we show the measured mean absolute PA discrepancy (hereafter, MAPAD)
between isophotal and re-Gaussianization shapes as a function of projected sky separation for
galaxies that have a bright $m_r < 19$ non-physically associated neighbor, as defined in (iii) of
Sec.~\ref{subsec:sample2} (plotted in purple open points). We see that the MAPAD increases to $\sim 27^\circ$
for the innermost sky separation bin, indicating potentially more contamination from bright neighbors at closer sky separation.

Besides systematics, there is also a noise contribution (see Sec.~\ref{sec: noise}) to the measured MAPAD, which must be estimated in order to properly constrain the level of bright neighbor and crowded field systematics.
Note that for the MAPAD data points shown in  purple open circles in Fig.~\ref{fig:PAdiff_skysep}, there is no contribution from a physical effect like isophotal twisting, due to our use of foreground/background galaxies at different redshifts from the bright central galaxies.

To distinguish between bright neighbor systematic and noise, we re-weight the galaxies in our
non-cluster field sample (as defined in Sec.~\ref{subsec:sample2}, (ii)) to match the distributions
of $m_r$, $z$, and ellipticity in the sample around bright galaxies used here.  This reweighting is
done separately within each bin in projected sky separation.  
We can then record the weighting factors in the $m_r$-$z$-ellipticity space, and use these weighting
factors to calculate the weighted-MAPAD from galaxies in the non-cluster field sample. The resulting
weighted-MAPAD value is then a proper estimation of the noise level for galaxies in each sky
separation bin, assuming that those three quantities are the main ones determining the statistical
uncertainty in the MAPAD.
 The triangular data points in both panels of Fig.~\ref{fig:PAdiff_skysep} show the resulting
 estimation of the noise level. The convergence of the triangular points towards the circular points
 at larger separations appears to validate the assumption behind this method.

\begin{figure*} 
\begin{center}
\includegraphics[trim=0.15cm 0cm 1.0cm 0cm,width=0.9\textwidth]{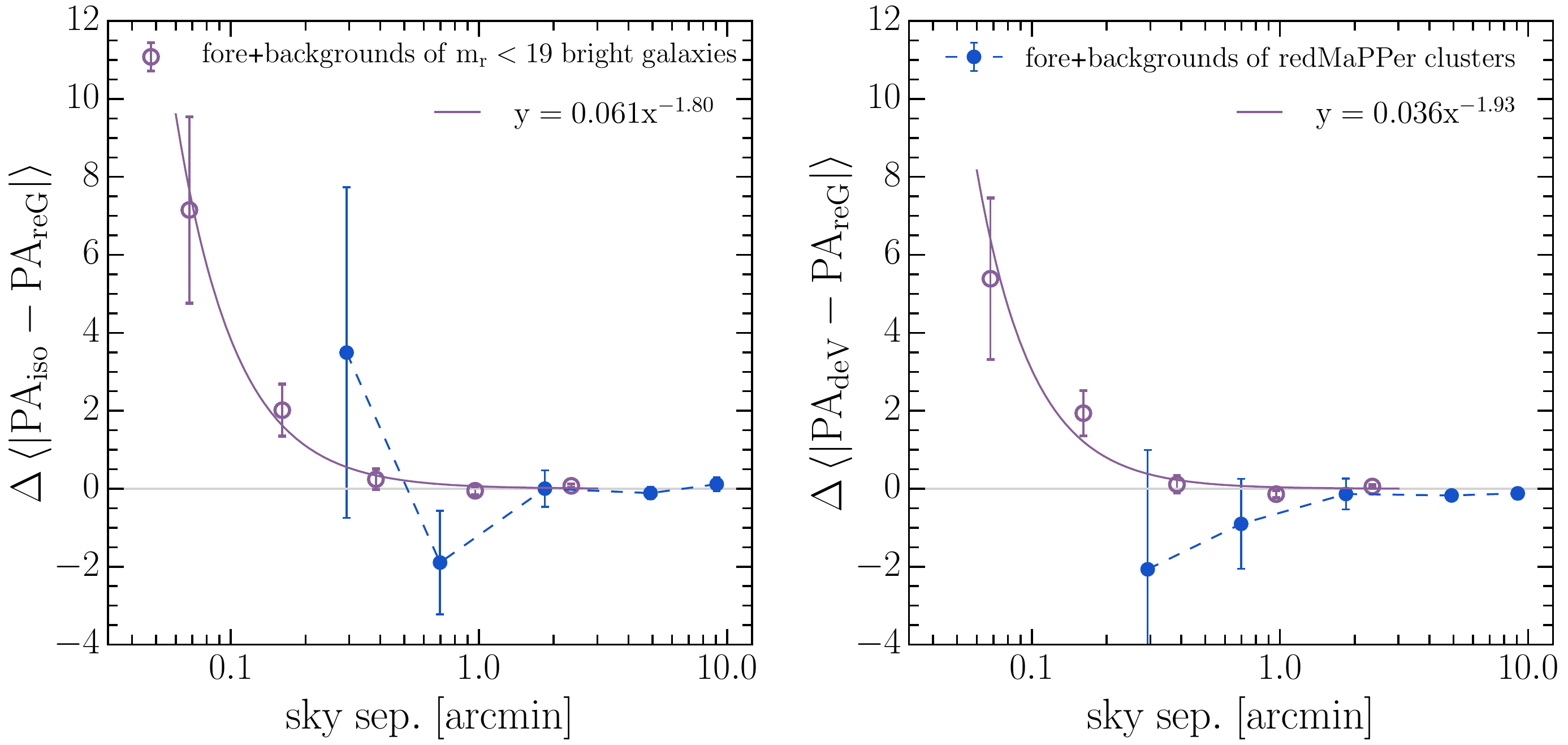}
\caption{Left panel: Excess of mean absolute PA differences (MAPAD) between isophotal and re-Gaussianization
  shapes caused by bright neighbor or crowded field systematics as a function of projected angular
  separation on the sky, for galaxies in the foreground and background of $m_r$ <19 (purple open circle) and 
  for galaxies in the foreground and background of redMaPPer (blue filled circle). Right panel: similar to the left panel, but for the excess of MAPAD
  between de Vaucouleurs and re-Gaussianization shapes. 
   \textit{As shown, the level of bright neighbor and crowded field systematics increases toward small sky
    separation bins.} The fitted models (see Eq.~\ref{eq:DPA_skysep}) between
  $\Delta\mean{\rm|PA_{iso/deV}-PA_{reG}|}$ and sky separation are provided in the legend with $x=$ sky
  separation measured in arcmin and
  $y=\rm{\Delta\mean{|PA_{iso/deV}-PA_{reG}|}}$.}
\label{fig:DeltaPAdiff_skysep}
\end{center}
\end{figure*}

After subtracting the noise contribution in the left panel of Fig.~\ref{fig:PAdiff_skysep}, the remaining signal shown in Fig.~\ref{fig:DeltaPAdiff_skysep} (purple open circles) should be dominated by MAPAD due to bright neighborhood systematics.  
This figure shows that for large sky separations ($\gtrsim$ 0.4$\arcmin$), the detected MAPAD is consistent with our prediction for the noise. For the
innermost bin in sky separation, the excess of MAPAD, $\rm{\Delta\mean{|PA_{iso}-PA_{reG}|}}$, increases to $\sim 7^\circ$ for
galaxies that have a bright ($m_r < 19$) neighbor. 
The best-fitting models of the form 
\begin{equation}\label{eq:DPA_skysep}
{ \rm{\Delta\mean{|PA_{iso/deV}-PA_{reG}|}}} = A\ {\rm{(sky\ separation)}}^{B}
\end{equation}
are provided in the legend of Fig.~\ref{fig:DeltaPAdiff_skysep}, and will later be used to estimate the
level of MAPAD due to bright neighbor systematics. 
The right panel of Fig.~\ref{fig:DeltaPAdiff_skysep} shows the same results as in the left panel,
but for de Vaucouleurs (rather than isophotal) vs.\ re-Gaussianization shapes. The trends in both panels are similar.

Aside from the simple diagnostics shown here, \citet{Mandelbaum05} and \citet{Aihara11} have also
pointed out other ways that the imperfect deblending and sky-subtraction can affect the measured
properties of the faint galaxy populations around bright galaxies in SDSS DR4 and DR8 photometry pipelines.
We conclude that bright neighbor systematic does play some role in the measured MAPAD
between different shape measurement methods in the
data, and its effect increases for galaxies around brighter neighbors.

We attempt to measure the strength of the crowded field systematic
 by measuring the MAPADs for foreground and background
galaxies of redMaPPer clusters as defined in sample (i) of Sec.~\ref{subsec:sample2}. The right
panel of Fig.~\ref{fig:PAdiff_skysep} shows the resulting estimate, with the estimated noise
contribution shown as well. 
The difference between the two sets of data points, as shown in the blue filled circles in Fig.~\ref{fig:DeltaPAdiff_skysep}, indicates the joint contribution of MAPAD due to both bright neighborhood and crowded field systematics in the redMaPPer cluster field.
Unfortunately here we lack pairs at small projected sky separation, resulting in a very noisy
estimate of these combined effects.

\subsection{PA discrepancies due to physical effects}
\label{sec: physical}

Finally, in the case that a pair of galaxies are physically associated, aside from noise and systematics, some portion of the measured MAPAD could be explained by a real physical effect known as ``isophote twisting'' \citep{diTullio78, Kormendy82}. 
The origin of this phenomenon is that the galaxy outer light profile may be more sensitive to
external tidal fields, and hence could show a stronger SA signal compared with its inner light profile. 

From $N$-body simulations, at one halo length scale, \citet{Pereira10} detected a
significant amount of isophotal twisting for triaxial galaxies orbiting in a cluster potential (see
their Figs.~6 and 8).
Also, \citet{Tenneti15} found in cosmological hydrodynamic simulations that the measured IA signal
at all spatial separations becomes larger when defining galaxy shapes and orientations in a way that
emphasizes the outer parts of the galaxy light profile.

Observationally, since the isophotal shape traces the outermost part of the light profile, the
measurement based on isophotal shapes should detect the strongest SA signals, followed by de
Vaucouleurs and re-Gaussianization shapes, if isophotal twisting is occurring at a significant
level. \citet{Singh16} detect a stronger IA amplitude with isophotal shapes than with
re-Gaussianization shapes at large separations ($\gtrsim$ 5Mpc). After considering possible
systematic errors, they conclude that this difference most likely originates from isophotal twisting.

In this work, since we focus particularly on galaxies in cluster environments, we need to reassess
whether systematics may be contributing in a significant way to the measured MAPADs between the two shape measurement methods compared to what was found in \citet{Singh16}.  Only after doing so can we draw conclusions about possible detections of isophotal twisting.

\begin{figure*}
\begin{center}
\includegraphics[trim=0.15cm 0cm 1.0cm 0cm,width=0.9\textwidth]{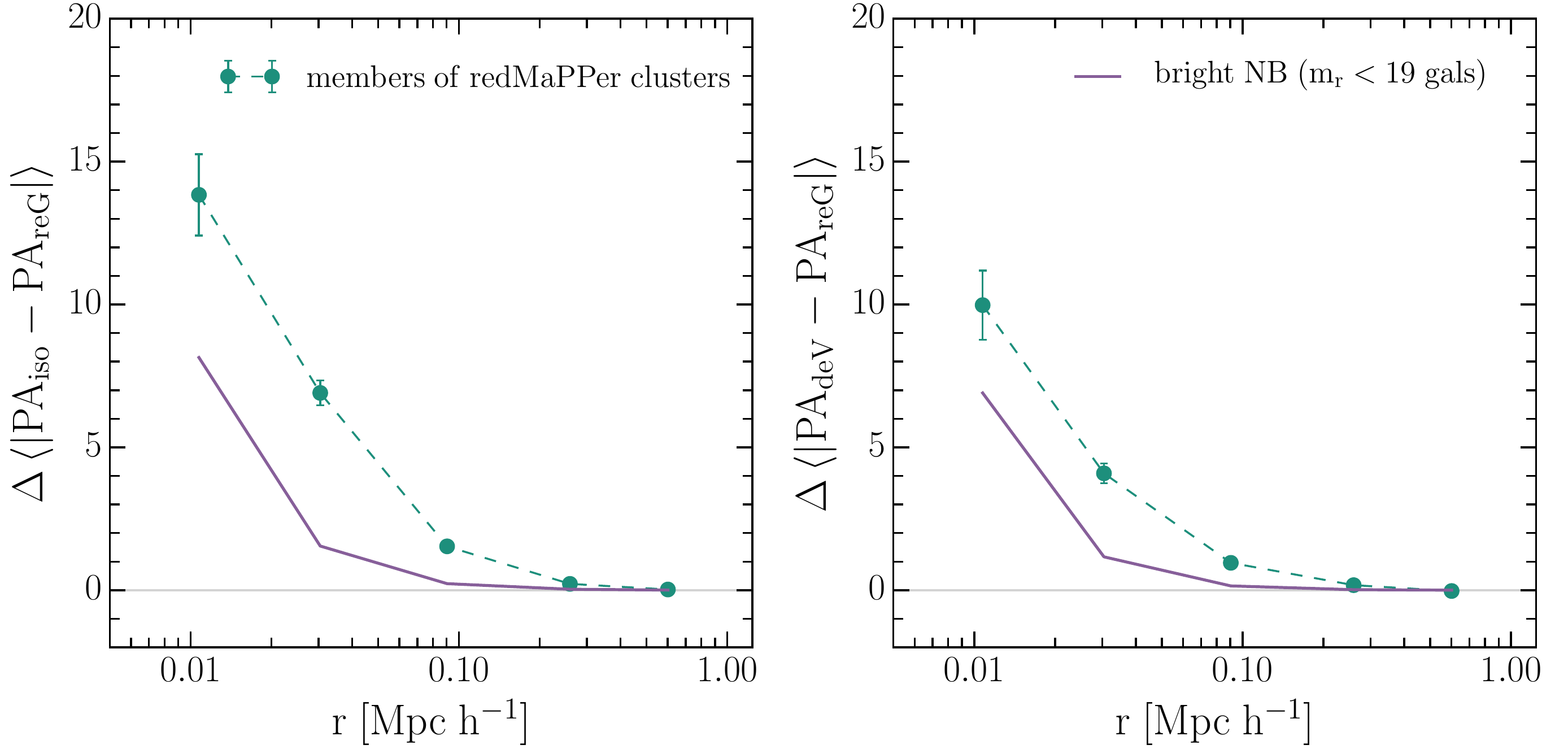}
\caption{Left panel: Excess of mean absolute PA differences (MAPAD) between isophotal and re-Gaussianization shapes, $\rm{\Delta{\mean{|PA_{iso}-PA_{reG}|}}}$, as a function of projected physical separation for members in redMaPPer clusters (teal green circles). Right panel: similar to the left panel but with $\rm{\Delta{\mean{|PA_{deV}-PA_{reG}|}}}$ in the $y$-axis.
The purple line is the estimated level of bright neighbor systematic derived using
the best-fitting models shown in Fig.~\ref{fig:DeltaPAdiff_skysep}. \textit{We observe that at the smallest $r$ bin, the bright neighbor systematic could potentially contribute more than 50\% of the measured $\rm{\Delta{\mean{|PA_{iso/deV}-PA_{reG}|}}}$.
Roughly, the remaining differences in the $y$-axes between the green dashed and purple solid lines are contributed by real physical 
isophote twisting signal and the crowded field systematic (for which we lack a good estimate), and probably 
some residual bright neighbor systematic due to the fact that the average apparent magnitude of cluster central galaxies is brighter than 
that of the galaxy sample used to estimate the bright neighbor systematic.
}
}
\label{fig:DeltaPAdiff_r}
\end{center}
\end{figure*}

The effect of isophote twisting should be more intense for galaxies in a stronger gravitational
field. Thus we expect to detect a larger MAPAD for satellites located physically closer, not just looked closer in projection, to the centers
of clusters. Therefore, in Fig.~\ref{fig:DeltaPAdiff_r}, we make similar plots for redMaPPer members (teal green circles) as that shown in Fig.~\ref{fig:DeltaPAdiff_skysep}, but change the x-axis from sky separation to physical projected separation. 
We process similar noise estimation as what we have done in Fig.~\ref{fig:PAdiff_skysep}, i.e. by reweighting non-cluster field galaxies to have similar $z$, $m_r$, and ellipticity distributions as that of redMaPPer members, but now operate it in bins of physical projected separation.
After the removal of noise, the values of y-axis shown in Fig.~\ref{fig:DeltaPAdiff_r} should be caused largely by bright neighbor and crowded field systematics, as well as physical isophote twisting effects.
The left panel of Fig.~\ref{fig:DeltaPAdiff_r} is the excess of MAPAD between isophotal and re-Gaussianization shapes, while the right panel plots that between de Vaucouleurs and re-Gaussianization shapes. 
In the following, we will first try to estimate how much of the excess MAPAD is contributed
  by the bright neighbor systematic, and then based on that we can judge whether there is leftover
  isophotal twisting signal.

To estimate what fraction of the excess MAPAD in Fig.~\ref{fig:DeltaPAdiff_r} is due to the bright neighbor systematic, 
we apply the following procedure: for each central-satellite pair of redMaPPer member, we use its projected sky separation as input in the 
derived best-fitting $\rm{\Delta\mean{|PA_{iso/deV}-PA_{reG}|}}$--$\rm{(sky\  separation)}$ relation (in the form of Eq.~\ref{eq:DPA_skysep}) shown in 
Fig.~\ref{fig:DeltaPAdiff_skysep} to estimate the level of systematics in cluster field. Next, we take averages for all central-satellite pairs in each physical $r$ bin.
The solid purple line in Fig.~\ref{fig:DeltaPAdiff_r} indicates the resulting estimated contribution due to bright neighbor systematics.
We observe that at the smallest $r$ bin, the bright neighbor systematic could potentially contribute
more than 50\% of the measured $\rm{\Delta{\mean{|PA_{iso/deV}-PA_{reG}|}}}$. However, the actual
level of bright neighbor systematic in cluster environments could be larger, since the central
galaxies have (on average) brighter apparent magnitudes than the sample of non-centrals used to estimate the bright neighbor systematic.

Roughly, the remaining differences in the $y$-axes between the green dashed and purple solid lines are contributed by real physical 
isophote twisting signal and the crowded field systematic (for which we lack a good estimate). 
Residual bright neighbor systematic error may also play a role since the average apparent magnitude of cluster central galaxies is brighter than 
that of the galaxy sample used to estimate the bright neighbor systematic.
Unfortunately, we have no good way to empirically estimate the crowded field systematic more accurately,
due to the small size of the foreground and background samples in cluster fields.
Simulation pipelines associated with future large surveys may be a better tool to estimate and
remove various sources of systematic contamination and provide constraints on isophote twisting
effects in cluster environments.


%% file: 6_discussion.tex
\section{The dependence of satellite alignment on the selected predictors}
\label{sec:discussion}

In Sec.~\ref{sec:LR}, we apply linear regression analysis and apply the model averaging technique to identify
predictors that have a significant influence on the variation of the SA signal (as summarized in Tables~\ref{tb:DR8reG_Mavg},~\ref{tb:DR7deV_Mavg} and ~\ref{tb:DR7iso_Mavg} for different shape measurements). We now address possible reasons for the observed relationship between $\phi_{\rm sat}$ and these selected predictors. 

\subsection{Dependence on satellite luminosity}
As shown in Figs.~\ref{fig:PhiSAT_x_DR8}a and~\ref{fig:PhiSAT_x_DR7}b, we found that satellite
$^{0.1}Mr$ has a very significant influence on the SA signal, with more luminous satellites being more likely to have their long axes oriented toward their host central galaxies. 

Our result is consistent with the observation of \citet{Hung12}. Based on high quality HST/ACS
data for shape measurement
they also detected a statistically significant trend for the dependence of $\mean{\phi_{\rm sat}}$ on
satellite luminosity (see their Fig.~6) for members in 12 X-ray clusters at 
$z\sim 0.5$--0.6. 

Based on $N$-body simulations, \citet{Pereira08} reported that there is no apparent dependence of 
subhalo alignment signals on the subhalo mass (see their Fig.~3). 
In a cosmological hydrodynamic simulation, \citet{Tenneti14} found that the misalignment between a
galaxy's own DM subhalo and its luminous component becomes larger for less massive
galaxies. Therefore, the observed relationship between $\mean{\phi_{\rm sat}}$ and satellite
luminosity may be due to this misalignment dependence on luminosity. For faint galaxies, which are
typically less massive, they appear to be more randomly oriented because their luminous components
are not good tracers of the orientation of their own DM halos.

\subsection{Dependence on satellite-central distance}
Satellite-central distance is another significant factor determining the strength of the SA effect,
with satellites located closer to their centrals having a stronger SA signal.  
Many previous observational studies that have reported detections of SA also found dependence of $\mean{\phi_{\rm sat}}$ on satellite-central distance \citep{Pereira05,Agustsson06,Faltenbacher07,Siverd09,Hung12}.

The satellite-central distance dependence naturally reflects the fact that SA is triggered by tidal
forces from the DM potential of the host halo. Hence, the strength of the tidal force would be
stronger for satellites located closer to the central region of the host halo. However, as discussed
in Sec.~\ref{sec: systematic}, part of this trend could coming from bright neighbor and crowded
field systematics, especially for galaxies located near the cluster central region. 

From the simulation side, where this radius-dependence can be measured to small scales without 
observational systematics, \citet{Pereira08} showed that the relationship between the subhalo SA signal and satellite-central distance is actually non-linear. 
As shown in their Fig.~4, the SA signal first rises gradually when satellites are closer to cluster
center, peaks at around 0.5 times the virial radius, then decreases again toward the center. This is
because when falling into the cluster along an eccentric orbit, a subhalo's orbiting speed becomes
too fast for the tidal torquing to be effective at the orbital pericenter, so the alignment cannot
keep up with the satellite subhalo's own motion (see Fig.~8 of \citealt{Pereira08}), leading to the decrease in SA signal at very small radius (see also discussion in Sec.~6 of \citealt{Kuhlen07}).

\subsection{Dependence on satellite ellipticity}
\begin{figure*}
\begin{center}
\includegraphics[trim=0.15cm 0cm 1.0cm 0cm,width=1\textwidth]{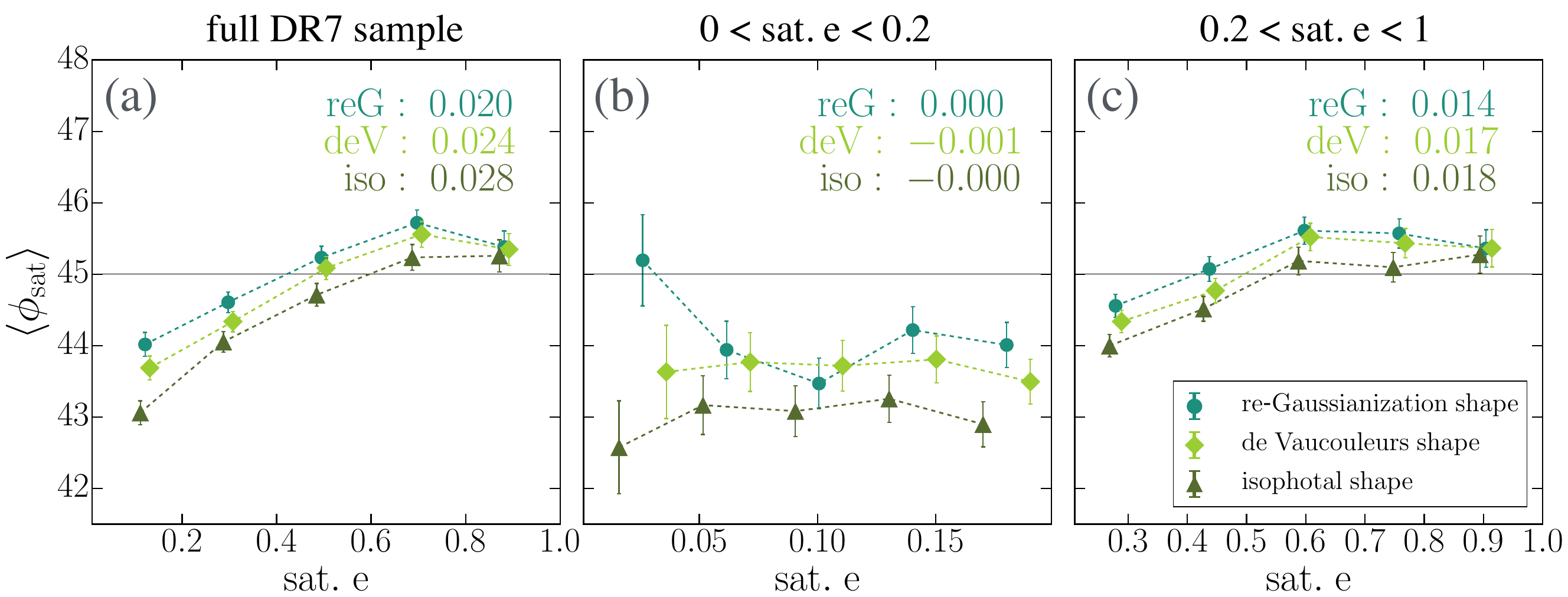}
\caption{The averaged SA angle, $\mean{\phi_{\rm sat}}$, in bins of satellite ellipticity. The left panel shows all satellites in our DR7 footprint sample; the middle panel shows only satellites with ellipticity $< 0.2$, while the right panel shows satellites with ellipticity $\ge 0.2$. 
The correlation coefficients between $\phi_{\rm sat}$ and satellite ellipticity measured in various shapes are provided in the upper right corner 
for each of subsamples.
\textit{We see the observed significant trend of positive correlation in the full satellite sample
  is still there for satellites with ellipticity $\ge 0.2$, whose PA (and thus $\phi_{\rm sat}$) measurements are more robust.}
}
\label{fig:check_sat_e}
\end{center}
\end{figure*}

We identified a statistically significant SA signal dependence on satellite ellipticity, with
rounder satellites exhibiting a stronger tendency to radially point toward their cluster central
galaxy. Since it is difficult to accurately determine the PA of a satellite with a round shape, we
divide our samples into two ellipticity bins at the boundary of 0.2, and see if the resulting
correlation is still strong enough that satellite ellipticity can be selected as an important
indicator of the SA effect through the variable selection process. Our result is that this quantity is still selected as a feature predictor. Fig.~\ref{fig:check_sat_e} plots the $\mean{\phi_{\rm sat}}$ value in bins of satellite ellipticity. 
We see that the net SA signal is fairly strong for satellites with ellipticity $< 0.2$ (Fig.~\ref{fig:check_sat_e}b), and that 
the detected positive correlation is still quite significant for satellites with ellipticity $\ge 0.2$ (Fig.~\ref{fig:check_sat_e}c).

According to \citet{Pereira10}, for galaxies orbiting in cluster potentials, their stellar
components tend to become more spherical with time (see their Fig.~18). Also it takes time for
satellites to become tidally locked and pointing radially toward central. We propose that the observed correlation between $\mean{\phi_{\rm sat}}$ and the ellipticity of satellites reflects this physical picture.

\subsection{Dependence on the $fracDeV$ parameter}
In the work of \citet{Siverd09}, based on a catalog of group-mass systems, they have studied the
effect of $fracDeV$, an indicator of a galaxy's bulge fraction, on the SA signal.  They reported that the level of SA strength is most strongly dependent on the $fracDeV$ parameter.
To compare with their result, we have added $fracDeV$ into our parameter pool for use during the variable selection process.

Our analysis also showed that $fracDeV$ is a statistically-significant predictor for the satellite
alignment effect, with de Vaucouleurs profile-dominated (higher bulge fraction) galaxies having stronger SA signal compared to galaxies with exponential dominated-profile (higher disk fraction). 
As shown in the $\phi_{\rm sat}$-$fracDeV$ plots of Figs.~\ref{fig:PhiSAT_x_DR8}d
and~\ref{fig:PhiSAT_x_DR7}e, the observed SA signal is mostly coming from satellites with high
$fracDeV$ values. 

\begin{figure*}
\begin{center}
\includegraphics[trim=0.15cm 0cm 1.0cm 0cm,width=1\textwidth]{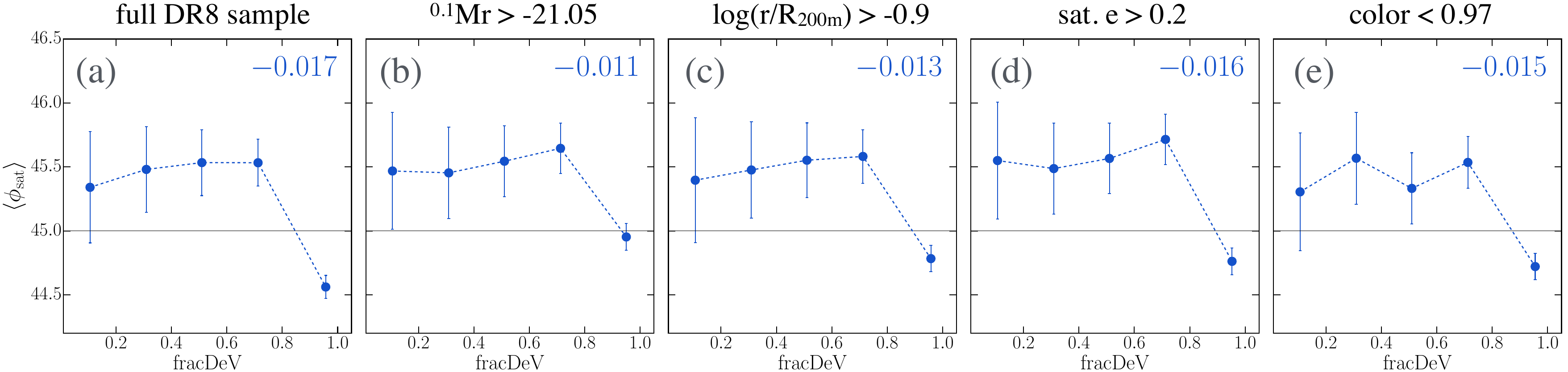}
\caption{The averaged SA angle, $\mean{\phi_{\rm sat}}$, in bins of $fracDeV$ for different
  subsamples of the redMaPPer satellites as indicated in the title of each plot. 
  The correlation coefficient between $\phi_{\rm sat}$ and $fracDeV$ for each subsample is shown in the upper right corner. 
    \textit{After
    excluding part of satellites with properties that correlate with $fracDeV$, the correlation
    trend between $fracDeV$ and $\phi_{\rm sat}$ still remains similar as the original full
    sample. This confirms that $fracDeV$ as its own distinct effect on the SA signal.}
}
\label{fig:check_fracDeV}
\end{center}
\end{figure*}

Typically a galaxy with higher luminosity, rounder shape, and redder color tends to have a higher bulge fraction, and thus higher $fracDeV$.
Therefore, it is important to ensure that the dependence between $\phi_{\rm sat}$ and $fracDeV$ is not caused by correlations of $fracDeV$ with other parameters. 
(Although linear regression is a useful tool to break out degeneracies among parameters, since SA is a weak signal, we still need to pay extra attention on the possible degeneracy issue.) 
To check whether $fracDeV$ is a representative parameter with its own distinct effect on
$\mean{\phi_{\rm sat}}$,  
we construct five subsamples by excluding satellites of the top 20\% most luminous, smallest
satellite-central distance, roundest, and reddest color from the parent DR8 sample pool each time.
After that, we can check if the remaining 80\% of the satellites still exhibit a significant correlation between $fracDeV$ and $\phi_{\rm
  sat}$.
In Fig.~\ref{fig:check_fracDeV}, we plot the averaged $\phi_{\rm sat}$ in bins of $fracDeV$
for these five subsample sets. One can see that although excluding these satellite subsets decreases the
overall SA signal strength, the trend between $fracDeV$ and $\mean{\phi_{\rm sat}}$ remains similar
as in the original full sample. This finding confirms that $fracDeV$ really is a special parameter
with its own distinct effect on the SA signal.

We propose that $fracDeV$ is chosen as an important predictor because it encapsulates information
about the importance of angular momentum in the dynamics of each galaxy. 
It has been observed that SA signal only appears in samples of red bulge-dominated galaxies, while
blue disky galaxies generally have random orientation within clusters 
(see e.g. \citealt{Pereira05,Faltenbacher07,Siverd09} based SDSS isophotal shapes and \citealt{Hung12} based on high quality HST images).
Since material in disks has higher angular momentum compared with that in bulges, it becomes less effective to torque disks to align with the surrounding tidal field.
According to the $N$-body simulation results of \citet{Pereira10}, galaxies with initial figure rotation generally take longer to become radially aligned than non-rotating galaxies (see their Sec~5.5). 
\citet{Tenneti16} also found using cosmological hydrodynamic simulations that the misalignment angle
between disky galaxies with the shape of their host DM subhalos is larger compared with
ellipticals.  In that case, they used an angular-momentum based discriminator for disk vs.\
elliptical galaxies, so again $fracDeV$ should be relevant.

We remind the readers that our work only includes fairly red galaxies, since the redMaPPer members are selected based on the red-sequence method. 
This may lead to the result that $fracDeV$ (tracing angular momentum) appears to be a more important
predictor than color (which directly reflects the gas contain of a galaxy). 
However, based on hydrodynamical simulations, \citet{Debattista15} pointed out that ``gas'' is a key factor affecting the degree of
misalignment between a disky galaxy and its own subhalo. 
A red, gas poor disk can have a stable orientation governed by halo torques, but when there is gas cooling onto a disk, the blue disk could have arbitrary orientations set by the balance between halo torques and angular momentum of the ongoing gas accretion.

\subsection{Dependence on central galaxy alignment angle $\theta_{\rm cen}$}
For de Vaucouleurs and isophotal shapes, we detected a positive correlation between $\phi_{\rm
sat}$ and $\theta_{\rm cen}$, with satellites located closer to the central galaxy major axis
direction showing a stronger SA signal.

In Paper I, we have explored the angular segregation of satellites in redMaPPer clusters and
concluded that the angular segregation may be due to 1) preferential infall of satellites along the filamentary structure aligned with the large-scale primordial tidal field (see Paper~I Sec.~6.1) or 2) the newly-established local tidal field produced by the current configuration of satellites which torques centrals to align (see Paper~I Sec.~7.3). 
The observed dependence of $\phi_{\rm sat}$ on $\theta_{\rm cen}$ can also be explained based on the above two scenarios.
Assuming that a central galaxy is aligned with its most dominant primordial tidal field, since many
satellites are fed into clusters along this direction, 
some satellites located near the edge of the cluster may still remember their original orientations
along this primordial tidal field because of their relatively late entrance into the clusters. Thus,
for satellites with small $\theta_{\rm cen}$, it is natural that they will point radially toward central galaxy \citep{Faltenbacher08}. 
For the second scenario, if later dynamical evolution has changed the central galaxy's orientation to align with its newly established local tidal field, satellites near cluster central region would also show tendency to align along this tidal field, especially those located at small $\theta_{\rm cen}$, forming a local filamentary structure.
Note it is likely that the later local tidal field still follows the direction with its primordial large-scale tidal field.

\begin{figure*}
\begin{center}
\includegraphics[trim=0.15cm 0cm 1.0cm 0cm,width=0.98\textwidth]{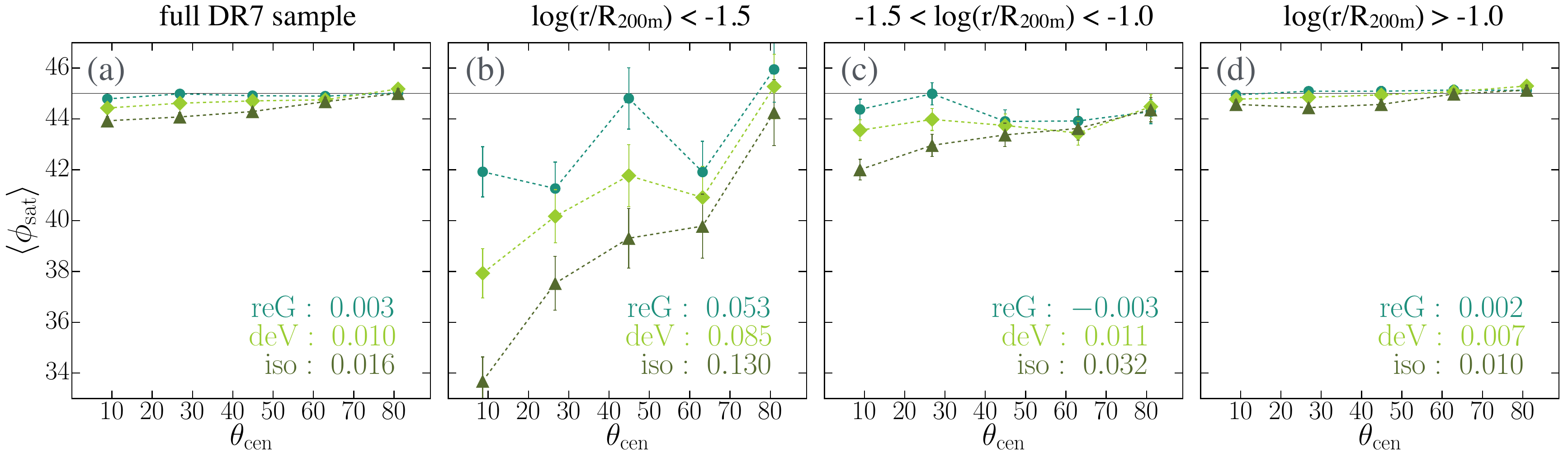}
\caption{The relationship between $\mean{\phi_{\rm sat}}$ and $\theta_{\rm cen}$ for the DR7 footprint sample in different satellite-central separation bins as indicated in the title of each plot. 
The correlation coefficients between $\phi_{\rm sat}$ and $\theta_{\rm cen}$ measured in various shapes are provided in the upper right corner 
for each of subsamples.
\textit{We see that the originally observed significant correlation in the full sample is dominated by satellites located closer to cluster central region.}
}
\label{fig:check_theta_cen}
\end{center}
\end{figure*}

One way to check the above scenarios is to look at the correlation between $\theta_{\rm cen}$ and $\phi_{\rm sat}$ for satellites at small and large satellite-central separations, as plotted in Fig.~\ref{fig:check_theta_cen}. We observe that there is almost no signal at the largest log($r$/R$\rm_{200m}$) bin (Fig.\ref{fig:check_theta_cen}d). The correlation is mostly driven by satellites at small log($r$/R$\rm_{200m}$) bin (Fig.\ref{fig:check_theta_cen}b). 
Thus if the detected correlation is real, then the local tidal field is the most likely cause for
the correlation between $\theta_{\rm cen}$ and $\phi_{\rm sat}$. 

Besides the physical origin, the correlation between $\theta_{\rm cen}$ and $\phi_{\rm sat}$ could be induce by systematics. 
At small satellite-central distances, $\phi_{\rm sat}$ measurements based on de Vaucouleurs and
isophotal shapes may suffer from bright neighbor systematic (see Sec.~\ref{sec: cluster field
  systematics}) due to the central galaxies' extended light profiles. Satellites located on the
major axes directions of the centrals would be more strongly affected by this systematic. 
This may be the reason why $\theta_{\rm cen}$ is identified as an important predictor only in de Vaucouleurs and isophotal shape measurements. 
It therefore remains interesting to check whether we can detect robust dependence between
$\theta_{\rm cen}$ and $\phi_{\rm sat}$, especially at small scale, using simulation
data in the future. 

\subsection{Dependence on redshift}

For the isophotal shape measurements, we observed that there is stronger SA for satellites at lower redshift.
As shown in Fig.~\ref{fig:PhiSAT_x_DR7}f, the correlation coefficient between $z$ and
$\mean{\phi_{\rm sat}}$ is very small ($0.002$), but the correlation is still identified using our variable selection procedure. 

However, we suspect that the correlation detected in isophotal shapes here may be dominated by
systematics. 
For a galaxy at lower redshift, its 25 mag/arcsec$^{2}$ isophote traces a larger area on the sky, and thus would have more fake alignment signal from bright neighbor and crowded field systematics (see Sec.~\ref{sec: cluster field systematics}). 
Besides, the correlation coefficients observed between $\phi_{\rm sat}$ and $z$ measured in
re-Gaussianization and de Vaucouleurs shapes are both negative, meaning that satellites at higher
redshift show stronger SA signal (although not a strong enough dependence to be identified through
our variable selection procedure).

Comparing with other observational work, \citet{Hao10} also observed an increase of the SA signal toward lower $z$ based on isophotal shape, but detected no SA signal across all redshift bins using de Vaucouleurs shapes. \citet{Schneider13} reported stronger SA at higher redshift for early-type galaxies in groups based on 2D Sersic model shape measurements.

From the simulation side, \citet{Pereira08} found that the SA strength increases steadily with time
for all of their simulated clusters (see their Fig.~5), suggesting that IA strength within the one-halo regime requires time to develop. 
This trend is inconsistent with the current best theoretical model for IA of galaxies at large scales. 
The linear alignment model \citep{Catelan01, Hirata04a} suggests that IA stems from the primordial tidal field at the time when galaxies form. 
This implies that later merging or baryonic processes of galaxy evolution may weaken this primordial
signal, as shown in the N-body simulation work of \citet{Hopkins05}, who found that the strength of
cluster alignments (not galaxy alignments within clusters) decreases at later times.

Currently, our technique is not good enough yet to completely demonstrate proper removal of systematics contamination
 and measure the evolution of SA signals with redshift robustly. 
Future exploration on the dependence of SA particularly with redshift is important in a sense that it may have 
different theoretical origin compared with current large-scale linear alignment model, which needs 
to be investigated to properly extend the linear alignment model down to smaller scale 
(see e.g. \citealt{Schneider10} for smaller scale IA modeling).


%% file: 7_summary.tex
\section{Summary}
\label{sec:summary}

In this work, we investigate the radial alignment of satellites in redMaPPer clusters based on three different shape measurement methods: re-Gaussianization, de Vaucouleurs and isophotal shapes.
We compare the observed SA signals among these measurements, and explore possible systematic effects.
To identify the predictors that are relevant to the variation of the SA signal, we perform linear regressions on all possible models and apply the model averaging technique on a total of 17 physical parameters related to satellite, central galaxy, and cluster properties (see Table~\ref{tb:predictors}), and quantify the significance of
their relationship with the satellite alignment angle, $\phi_{\rm sat}$.

Our main results are summarized as follows: 

\begin{enumerate}
\item Based on re-Gaussianization shape measurements, which puts more weight on a galaxy's inner
  light profile, we do not detect any convincing SA signal in the overall $p_{\rm mem}>0.55$ satellite population of redMaPPer clusters (Fig.~\ref{fig:phisat_dist_DR8}).
However, a statistically significant SA signal is observed for the entire sample when using de
Vaucouleurs shapes, and the overall SA strength reaches to its strongest level in isophotal shape,
which traces the outermost light profile of a galaxy (Fig.~\ref{fig:phisat_dist_DR7}). 

\item Despite the lack of detection of satellite alignments for the entire sample, there are
  nonetheless distinct subpopulations that carry highly significant satellite alignments signals
  when measuring using re-Gaussianization. The SA strength is strongest for satellites with higher
  luminosity, located closer to their central galaxies, with smaller ellipticity, and have higher bulge
  fraction in the light profiles.
  (Fig.~\ref{fig:PhiSAT_x_DR8}). We also find that satellites located
  closer to the major axis directions of their central galaxies show higher SA signal, when using de
  Vaucouleurs and isophotal shape measurements for satellite shapes (Fig.~\ref{fig:PhiSAT_x_DR7}). 

\item The selected predictors that show significant influence on the SA effect highlight the roles of tidal torquing mechanism, the primordial pre-infall alignment, the process of violent relaxation in clusters, and the angular momentum of galaxies in causing the observed SA dependences. 

\item We discuss possible factors that could cause the observed different strength of SA among these three shape measurement methods (Fig.~\ref{fig:PA_difference}),
provide an estimate of the noise level and discuss contributions from systematics and physical isophote twisting effect.

\end{enumerate}

Over the past decade, there has been some disagreement in the literature regarding the existence and
strength of the SA effect. 
Here we report detections of SA phenomenon based on a well-understood shape measurement method using
nearly $10^4$ clusters, which provides great statistical power in constraining SA.
We identify the regions of parameter space where SA signals become significant, which in some cases
can explain previous reported non-detections (e.g., in measurements dominated by satellites in the
regions of parameter space where we also find no significant detection of SA).  
Our results will be useful in improving IA modeling at small scale, for example by building a more
realistic halo model of intrinsic alignments \citep[building on work by][]{Schneider10}, and further
helps constrain systematics from IA in weak lensing analysis. 
We also discuss possible physical origins of the SA signal based on the galaxy and cluster
properties that most strongly predict it, and point out directions for future work with even larger
cluster samples that are becoming available with next-generation imaging surveys.


%% file: ack.tex
We thank Eric Baxter for providing galaxy concentration parameter used in the work, and 
Sukhdeep Singh for helpful comments and discussions. 
We also thank Sivaraman Balakrishnan for sharing his view on various model selection criteria. 
This work was supported by the National Science Foundation under
Grant No.\ AST-1313169 and by NASA ROSES 12-EUCLID12-0004.

%% file: 8_appendix.tex

\section{The choice of membership probability cut of our sample}
\label{app:Pmem cut}

There are several considerations driving the choice of $p\rm_{mem}$ cut when defining the sample to
use for measuring SA.  In general, a higher $p\rm_{mem}$ cut would result in a
stronger SA signal because brighter and redder satellites tend to have a higher $p\rm_{mem}$ in
redMaPPer, and these galaxies are more likely to point radially toward cluster centers. 
However, raising the threshold in $p\rm_{mem}$ results in a smaller sample size, and thus
larger statistical uncertainty. Setting a lower $p\rm_{mem}$ cut would increase the sample size
(reduce statistical errors), but also results in a lower signal due to both the inclusion of lower
luminosity satellites which carry less signal, and due to the higher contamination rate from
non-cluster members. In this appendix,
we attempt to estimate the SA signal and noise as a function of the $p\rm_{mem}$ cut.
Based on the observed dependences, we can then determine a $p\rm_{mem}$ cut 
that maximizes the S/N of SA.


Here we define the signal $S$ to be the weighted averaged SA angle over all
potentail central-satellite pairs indexed $i$, 
\begin{equation} \label{eq:S1}
S = \mean{45^{\circ}-\phi_{\rm sat}} = \frac{\sum\limits_{i} w_i (45^{\circ}-\phi_{{\rm sat}, i})}{\sum\limits_{i} w_i},
\end{equation}
where $w_i$ is the weighting factor for each central-satellite pair.  In practice, we use  $w_i=p_{{\rm mem},i}$. We shift $\phi_{\rm sat}$ by 45$^{\circ}$ such that for pairs with no alignment, their signal $S=0$.
Some of the central-satellite pairs used in our calculation may be contaminated by
foreground/background galaxies that are not physically associated with the cluster system. These
fake pairs are assumed to contribute nothing to the numerator of Eq.~\eqref{eq:S1}; in other words,
we are ignoring the lensing of background galaxies, which would lead to
$\mean{\phi_\text{sat}}>45^{\circ}$.  We can rewrite Eq.~\eqref{eq:S1} as
\begin{equation} \label{eq:S2}
S = \frac{\sum\limits_{{\rm real\ pair,}j} w_j (45^{\circ}-\phi_{{\rm sat}, j})}{\sum\limits_{i} w_i},
\end{equation}
with the summation in the numerator (indexed $j$) now including only contributions from real
central-satellite pairs. The contaminating pairs do contribute to the denominator, thereby diluting
the signal. 
Under the assumption that the $p_{\rm mem}$ values in the redMaPPer catalogue represent a correct
statistical description of cluster membership \citep[but see][]{2016arXiv161100366Z}, the summation
$\sum\limits_{{\rm real\ pair},j}$ should be statistically equivalent to $\sum\limits_{i} p_{{\rm
    mem},i}$. 
 Also, 
 we find that 
 a linear model is a good description 
 for the relationship between the strength of the SA signal $(45^{\circ}-\phi_{{\rm sat}})$
 and $p_{\rm mem}$ \footnote{We have tried fitting the relation between $(45^{\circ}-\phi_{{\rm sat}})$ and $p_{\rm mem}$ with linear, quadratic, and other higher order of power law models. Unfortunately, our data does not have the power to constrain more complicated models very well, so we simply apply the linear model here.}.


With these two simplifications and also putting
$w_i=p_{{\rm mem},i}$ in Eq.~\eqref{eq:S2}, we conclude that the signal $S$
is described as 
\begin{equation} \label{eq:S3}
S \propto \frac{\sum\limits_{i} p_{{\rm mem},i}^3}{\sum\limits_{i} p_{{\rm mem},i}}.
\end{equation}
The first $p_{{\rm mem},i}$ in the numerator comes from replacing the summation over real
central-satellite pairs with the summation over all potential pairs; the second comes from the
weight; and the third from the approximation that individual SA signal to lowest
order are proportional to $p_{{\rm mem},i}$

For the noise N, we simply apply standard error of the mean as the level of noise here, i.e. $N = \sigma_{\phi_{\rm sat}} / \sqrt{N_{\rm eff}}$.
Here the standard deviation of $\phi_{\rm sat}$ ($\sigma_{\phi_{\rm sat}}$) is around 26$^\circ$ per
$p_{\rm mem}$ bin, while the the effective number of central-satellite pairs ($N_{\rm eff}$) is
ranging from $10^4$ to $5 \times 10^4$ across various $p_{\rm mem}$ bins. Therefore, the variation
in the noise on the measurement of the average $\phi_{\rm sat}$ value with the $p_{\rm mem}$ cut
value is mostly driven by the change in $N_{\rm eff}$. We assume that the noise $N$ is proportional to the Poisson noise on the mean value, defined as
\begin{equation}
N \propto \frac{1}{\sqrt{N_{\rm eff}}} = \frac{1}{\sqrt{\sum\limits_{i} p_{{\rm mem},i}}}
\end{equation} 

\begin{figure} 
\begin{center}
\includegraphics[width=0.45\textwidth]{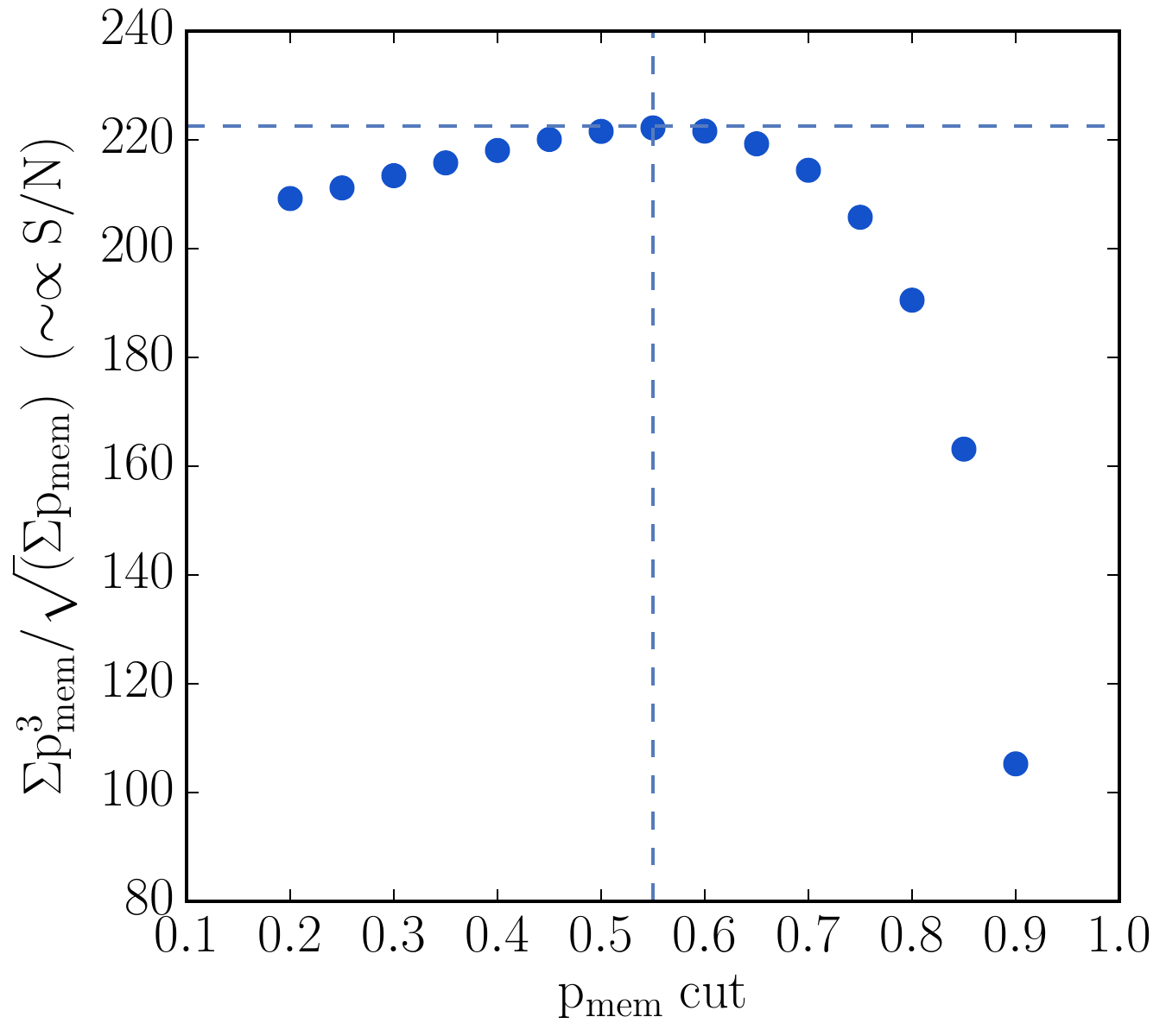}
\caption{Satellite alignment $S/N$ optimization with different $p_{\rm mem}$ thresholds. The $S/N$
  is maximized at $p_{\rm mem}$ cut = 0.55. Note that the absolute normalization of the curve is
  arbitrary; only the relative changes as a function of the $p_{\rm mem}$ cut are meaningful.}
\label{fig:pmem_cut}
\end{center}
\end{figure}

The signal to noise ratio $S/N$ would then be proportional to:
\begin{equation} \label{eq:S/N}
S/N \propto \frac{\sum\limits_{i} p_{{\rm mem},i}^3}{\sqrt{\sum\limits_{i} p_{{\rm mem},i}}}.
\end{equation}
This is a quantity that we can easily calculate for all central-satellite pairs in our catalog, as a
function of the lower limit on $p_{\rm mem}$.  The results are shown in Fig.~\ref{fig:pmem_cut}. 
A $p_{\rm mem}$ cut at 0.55 gives the highest $S/N$, so we adopt this value throughout our analysis.

%% file: 9_appendix2.tex

\section{Estimating $\mean{\phi_{\rm sat}}$ with contamination from background lensing signal}
\label{app:est_len}

In Fig.~\ref{fig:PhiSAT_x_DR8}, we saw that there are some subsamples with measured $\mean{\phi_{\rm sat}}>45^\circ$, meaning that satellites in these data bins show a net preferred tangential alignment signal. 
Part of the excess may come from lensing of galaxies that are actually in the background, because we
allow a $p_{\rm mem}$ cut at 0.55 when doing our analysis. 
Here we provide a rough estimate of the degree of this contamination 
based on the re-Gaussianization shape results
\footnote{In principle, lensing contamination comes into all shape measurements, but 
for de Vaucouleurs and isophotal shapes, it is hard to extract the hidden lensing signal due to higher levels of systematics in these shape measurements.}.

We begin by defining two quantities: fraction of expected contamination from fake members ($f_x$)
and fraction of expected contribution from true members ($f_v$).  Assuming that the $p_{\rm mem}$
values provide a correct statistical reflection of reality, we can estimate these via summations
over all central-satellite pairs with $p_{\rm mem}>0.55$, indexed $i$:
\begin{equation}\label{eq:fx}
f_x = \frac{\sum\limits_{i} w_i (1-p_{\rm mem})}{\sum\limits_{i} w_i}
\end{equation}
\begin{equation}\label{eq:fv}
f_v = \frac{\sum\limits_{i} w_i (p_{\rm mem})}{\sum\limits_{i} w_i}
\end{equation}
Here the weighting factor is just $p_{\rm mem}$, and 
by definition, $f_x+f_v=1$. 

After $f_x$ and $f_v$ are calculated, we can estimate the predicted value of $\mean{\phi_{\rm sat}}$
due to background contamination by: 
\begin{equation}\label{eq:est_phisat}
\mean{\phi_{\rm sat}}_{\rm pred}= f_v \times 44.92 + f_{\rm back} f_x \times 46.07 + (1-f_{\rm back}) f_x \times 45.0
\end{equation}
Here $f_{\rm back}$ is a free parameter that controls the fraction of contaminating pairs for which
the contaminating galaxy is in the background (while $1-f_{\rm back}$ is the fraction of
contaminating pairs consisting of foregrounds). 
The predicted $\mean{\phi_{\rm sat}}$ value from each component is taken from that shown in the legend of Fig.~\ref{fig:phisat_dist_DR8}. 
For real cluster members, $\mean{\phi_{\rm sat}}=44.92$; for background pairs $\mean{\phi_{\rm sat}}=46.07$, while for foregrounds, we expect $\phi_{\rm sat}$ to have $\mean{\phi_{\rm sat}}=45.0$.
The triangular orange data points shown in Fig.~\ref{fig:PhiSAT_x_DR8} are estimated based on
setting $f_{\rm back}=0.6$.
Although here we do not have a good estimation for $f_{\rm back}$, we find that the derived $\mean{\phi_{\rm sat}}_{\rm pred}$ are all roughly around 45.1$^{\circ}$, insensitive to the setting of $f_{\rm back}$ values ranging from 0.5$\sim$1.
